\documentclass[10pt,prd,superscriptaddress,twocolumn,nofootinbib]{revtex4-2}
\usepackage[paper=letterpaper,margin=1in]{geometry}
\usepackage[utf8]{inputenc}
\usepackage{dsfont}
\usepackage[dvipsnames]{xcolor}
\usepackage{amsfonts,amsmath,amssymb}
\allowdisplaybreaks[4]
\usepackage{euscript}
\usepackage{hhline}

\usepackage{braket}
\usepackage{amsmath}
\DeclareMathOperator{\arctanh}{arctanh}
\usepackage{starfont}
\usepackage{color,soul}  

\usepackage{tensor}
\usepackage{amsthm}
\usepackage{graphicx}
\usepackage{slashed}
\usepackage{leftidx}
\usepackage{subfigure}
\usepackage{bm}
\usepackage{bbm}
\usepackage{enumitem}
\definecolor{outerspace}{rgb}{0.25, 0.29, 0.3}
\definecolor{scarlet}{rgb}{1.0, 0.13, 0.0}
\usepackage[header,title,page,titletoc]{appendix}  
\definecolor{princetonorange}{rgb}{1.0, 0.56, 0.0}
\definecolor{WildStrawberry}{rgb}{1.0, 0.26, 0.64}
\definecolor{rossocorsa}{rgb}{0.83, 0.0, 0.0}
\definecolor{navyblue}{rgb}{0.0, 0.0, 0.5}
\usepackage{comment}
\usepackage{cancel}
\usepackage[normalem]{ulem}

\usepackage{mathtools}

\newcommand{\dd}{{\rm d}}

\makeatletter
\newcommand{\dal}{\mathop{\mathpalette\dal@\relax}}
\newcommand{\dal@}[2]{%
  \begingroup
  \sbox\z@{$\m@th#1\square$}%
  \dimen0=\fontdimen8
    \ifx#1\displaystyle\textfont\else
    \ifx#1\textstyle\textfont\else
    \ifx#1\scriptstyle\scriptfont\else
    \scriptscriptfont\fi\fi\fi3
  \makebox[\wd\z@]{%
    \hbox to \ht\z@{%
      \vrule width \dimen0
      \kern-\dimen0
      \vbox to \ht\z@{
        \hrule height \dimen0 width \ht\z@
        \vss
        \hrule height 2\dimen0
      }%
      \kern-2.5\dimen0
      \vrule width 2.5\dimen0
    }%
  }%
  \endgroup
}

\makeatother

\usepackage{multirow}
\usepackage{changepage}
\usepackage{upgreek}

\newcommand{\dif}{\mathrm{d}} 
\newcommand{\M}{\mathsf{M}} 
\newcommand{\p}{\mathsf{p}} 
\newcommand{\N}{\mathrm{N}} 
\newcommand{\GN}{G_\mathrm{N}} 
\newcommand{\nmax}{{n_\mathrm{max}}} 
\newcommand{\vrho}{\varrho} 

\newlength{\Sone}
\newlength{\Stwo}
\newlength{\Sthree}
\newlength{\Sfour}
\newlength{\Sonee}
\newlength{\Stwoo}
\newlength{\Sthreee}
\newlength{\Sfourr}
\newcommand{\be}{\begin{equation}}
\newcommand{\ee}{\end{equation}}

\usepackage{wasysym}
\usepackage{hyperref}
\hypersetup{
    colorlinks,
    citecolor=Peach,
    filecolor=Peach,
    linkcolor=MidnightBlue,
    urlcolor=MidnightBlue
}

\begin{document}

\title{Regular Geometries from Singular Matter in Quasi-Topological Gravity}

\author{Pablo Bueno}
\email{pablobueno@ub.edu}
\affiliation{Departament de F\'isica Qu\`antica i Astrof\'isica, Institut de Ci\`encies del Cosmos\\
 Universitat de Barcelona, Mart\'i i Franqu\`es 1, E-08028 Barcelona, Spain}

\author{Robie A. Hennigar}
\email{robie.a.hennigar@durham.ac.uk}
\affiliation{Centre for Particle Theory, Department of Mathematical Sciences,\\ Durham University, Durham DH1 3LE, UK}

\author{\'Angel J. Murcia}
\email{angelmurcia@icc.ub.edu}
\affiliation{Departament de F\'isica Qu\`antica i Astrof\'isica, Institut de Ci\`encies del Cosmos\\
 Universitat de Barcelona, Mart\'i i Franqu\`es 1, E-08028 Barcelona, Spain}

\author{Aitor Vicente-Cano}
\email{avicentecano@icc.ub.edu}
\affiliation{Departament de F\'isica Qu\`antica i Astrof\'isica, Institut de Ci\`encies del Cosmos\\
 Universitat de Barcelona, Mart\'i i Franqu\`es 1, E-08028 Barcelona, Spain}


\begin{abstract}
Vacuum quasi-topological gravity with infinitely many terms in the action satisfies Markov's limiting curvature hypothesis: the spherically symmetric solutions are regular and all curvature invariants are bounded by solution-independent scales. We study how this picture changes when the theory is coupled to matter. We find that minimally coupled matter spoils the scaling properties of the vacuum equations that lead to the validity of Markov’s hypothesis but find the corresponding geometries often remain regular. We make this precise by developing a set of sufficient conditions on general static, spherically symmetric stress-tensors such that the corresponding solutions have bounded curvature. These conditions cover regular matter sectors but also singular matter profiles that are {\it sufficiently singular} in a sense we quantify. Our conclusions hold independently of the matter field equations and include configurations in which matter exhibits divergent energy density and pressure at finite radius or at Killing horizons, results that may have implications for mass inflation in these models. We then explore non-minimal couplings, focusing on theories with infinite towers of higher-curvature and electromagnetic terms in the action. In this class, Markov’s hypothesis can be restored: we present theories admitting a universal upper bound on curvature, independent of the mass and charge. Overall, our results highlight subtleties in coupling quasi-topological gravity to matter and suggest Markov’s hypothesis as a potential selection criterion for resummed gravity–matter effective theories.

\end{abstract}
\onecolumngrid
\maketitle

\twocolumngrid

\vspace{-.3cm}

\section{Introduction}
According to general relativity (GR), black holes possess spacetime singularities in their interiors~\cite{Hawking:1973uf,Senovilla:1998oua}. As experimental evidence for the existence of black holes~\cite{LIGOScientific:2016aoc,LIGOScientific:2025slb,LIGOScientific:2025rid,EventHorizonTelescope:2019dse}---or objects that closely resemble them~\cite{Cardoso:2019rvt}---continues to accumulate, the theoretical question concerning the fate of singularities becomes increasingly pressing, despite their experimental inaccessibility.

Among the possible alternatives, a conceptually conservative yet physically radical approach is provided by regular black holes (RBHs): spacetimes which preserve an exterior event horizon while replacing the central singularity with a regular core concealed behind an inner horizon~\cite{Sakharov:1966aja, 1968qtr..conf...87B, Poisson:1988wc,Dymnikova:1992ux,Hayward:2005gi}. Remarkable progress regarding the physical properties of RBHs has been achieved relying on purely kinematical considerations~\cite{Carballo-Rubio:2018pmi,Carballo-Rubio:2021bpr, Carballo-Rubio:2022kad,Carballo-Rubio:2025fnc,LimaJunior:2025uyj}. However, the absence of a general dynamical framework has remained the Achilles’ heel of the RBHs program since its inception more than half a century ago.  

To date, the most widely investigated approach to realizing RBHs as solutions of concrete theories has been to couple GR to suitably chosen---more or less exotic---matter~\cite{Ayon-Beato:1998hmi,Bronnikov:2000vy,Ayon-Beato:2000mjt,Bronnikov:2000yz,Ayon-Beato:2004ywd,Dymnikova:2004zc,Berej:2006cc,Balart:2014jia,Fan:2016rih,Bronnikov:2017sgg,Junior:2023ixh,Alencar:2024yvh,Bronnikov:2024izh,Bolokhov:2024sdy,Skvortsova:2024wly,Murk:2024nod,Li:2024rbw,Zhang:2024ljd}. Nevertheless, generic black hole solutions in these models remain singular, with RBHs occurring only in highly restricted corners of parameter space. In particular, since all vacuum solutions of GR persist as solutions, the underlying singularity problem remains essentially unaltered. Alternative routes and mechanisms have been proposed \emph{e.g.,} in~\cite{Frolov:1989pf,Barrabes:1995nk,Bonanno:2000ep,Nicolini:2005vd,Greenwood:2008ht,Olmo:2012nx,Saini:2014qpa,Frolov:2014jva,Balakin:2015gpq,Bazeia:2015uia,Chamseddine:2016ktu,Bambi:2016xme,Bejarano:2017fgz,Colleaux:2017ibe,Kawai:2017txu,Cano:2018aod,Colleaux:2019ckh,Guerrero:2020uhn,Cano:2020qhy,Cano:2020ezi,Brandenberger:2021jqs,Bueno:2021krl,Olmo:2022cui,Junior:2024xmm,Harada:2025cwd,Bueno:2025dqk}. 

Recently, static RBHs corresponding to multiparametric generalizations of the Schwarzschild solution in general dimensions have been shown to arise as the unique spherically symmetric solutions of broad classes of theories involving GR coupled to infinite towers of higher-curvature corrections~\cite{Bueno:2024dgm}. All such theories belong to the so-called \emph{Quasi-topological} (QT) class~\cite{Oliva:2010eb,Quasi,Dehghani:2011vu,Ahmed:2017jod,Cisterna:2017umf,Bueno:2019ycr,Bueno:2022res,Moreno:2023rfl,Moreno:2023arp}, characterized by possessing second-order equations on spherically symmetric backgrounds, as well as for satisfying versions of the Birkhoff theorem~\cite{Bueno:2025qjk}. With the exception of the $D=4$ case---which requires non-polynomial densities~\cite{Bueno:2025zaj}\footnote{See also~\cite{Chinaglia:2017wim, Colleaux:2017ibe,Colleaux:2019ckh} for previous results involving RBHs as solutions to non-polynomial gravities.}---the theories are perturbatively equivalent, via field redefinitions, to the most general gravitational effective action~\cite{Bueno:2019ltp,Bueno:2024dgm}.
Remarkably, within this framework such RBHs arise as the byproduct of the gravitational collapse of ordinary matter~\cite{Bueno:2024zsx,Bueno:2024eig,Bueno:2025gjg}.  Thus, QT theories offer a general dynamical framework within which a wide range of questions concerning the physical properties of RBHs can be systematically probed for the first time.\footnote{For additional progress on the properties of RBHs in QT models, see~\cite{Konoplya:2024kih,Konoplya:2024hfg,DiFilippo:2024mwm,Ma:2024olw,Cisterna:2024ksz,Ditta:2024iky,Wang:2024zlq,Hennigar:2020kqt,Fernandes:2025eoc,Fernandes:2025fnz,Cisterna:2025vxk,Ling:2025ncw,Eichhorn:2025pgy,Fernandes:2025mic,Aguayo:2025xfi,Boyanov:2025pes,Konoplya:2025uta,Hao:2025utc,Tan:2025hht,Frolov:2025ddw,Chen:2025iuy,Arbelaez:2026eaz,Li:2026mam,Borissova:2026wmn,Tsuda:2026xjc,Konoplya:2026gim,Senovilla:2026fby,Borissova:2026krh}.} In this paper we address one such question, namely: how does singular matter affect the regularity of spacetimes that are non-singular in the vacuum case? 

As shown in~\cite{Frolov:2024hhe}, (vacuum) RBHs in any QT theory satisfy a version of Markov's limiting curvature hypothesis (LCH)~\cite{Frolov:1989pf,PismaZhETF.36.214}, \emph{i.e.,} each of their curvature invariants remains bounded above by a certain universal mass-independent quantity.\footnote{This is a highly non-trivial property which many proposed RBH models, such as the Bardeen black hole~\cite{1968qtr..conf...87B}, fail to satisfy.} The situation changes in the presence of matter. Indeed, as shown in~\cite{Bueno:2025tli}, perfect-fluid stars in QT theories may reach arbitrarily high curvatures as generalized versions of the Buchdahl limit are approached, unless the dominant energy condition is imposed. Nevertheless, the interplay between singular matter and regular geometry turns out to be remarkably more subtle and counterintuitive in general. Indeed, in this paper we perform a general analysis of this interplay for minimally coupled, static, spherically symmetric matter, remaining largely agnostic about the details of the matter model. We observe that regular matter sectors produce regular spacetimes, and so do matter configurations with central power-law density singularities. On the other hand, spacetime singularities do appear in the presence of matter with mildly singular energy densities at finite radii. Interestingly, the trend is reversed for sufficiently singular matter. In that case, which includes configurations with divergent energy density and pressure at finite radii or at Killing horizons, the regularity of spacetime can be preserved.

Our results align with a recent striking finding concerning the fate of mass inflation at the inner horizon of RBHs in this setting~\cite{Frolov:2026rcm}. Specifically, by analyzing a model of two null shells colliding near the inner horizon, Frolov and Zelnikov demonstrated that the mass inflation instability does not develop for RBHs in QT theories, at least within the regime where a classical metric description remains valid.  

In another vein, understanding the actual interplay of matter and geometry in general effective theories requires the consideration of higher-order terms in curvature and matter, including non-minimal couplings. As a first step in this program, we also study special theories of gravity and a gauge $(D-3)$-form which admit magnetically charged black hole solutions characterized by a single metric function fulfilling an algebraic equation. Such~\emph{electromagnetic quasi-topological gravities}~\cite{Cano:2020ezi,Cano:2020qhy,Bueno:2021krl,Cano:2022ord,Bueno:2022ewf,Bueno:2022jbl,Bueno:2025dqk} have been identified in general space-time dimensions $D\geq 3$ and may incorporate non-minimally coupled terms of any order in the curvature and gauge field strength. We will discuss in which cases these theories admit magnetically charged regular black holes, as well as specify sufficient conditions for Markov's limiting curvature hypothesis to hold in electrovacuum solutions.

The structure of the remainder of the paper is the following. In Section~\ref{sssm} we explain the general setup which we exploit throughout most of the paper. This entails considering general static and spherically symmetric solutions of QT gravities coupled to a minimally coupled---but otherwise general---matter stress tensor. We obtain the general form of all possible combinations of the metric functions susceptible of appearing in general algebraic curvature invariants. In Section~\ref{lchv} we review the argument which establishes the fulfillment of Markov's limiting curvature hypothesis for general vacuum spherically symmetric regular black holes in QT theories, as well as the conditions for such solutions to exist.
In Section~\ref{mstt} we analyse the interplay between singular matter and regularity in the case of matter satisfying $T_t^t=T_r^r$. We show that singular matter leads to singular geometries whenever the singularity is mild enough, whereas regular geometries are produced by sufficiently singular matter.
In Section~\ref{gc} we consider general stress tensors. We find a similar pattern to the one observed in the restricted case of the previous section.  In particular, we show that inner and cosmological horizons remain regular in the presence of matter densities and radial pressures which blow up at such horizons.
Some explicit examples corresponding to minimally coupled electromagnetic and scalar fields are worked out in Section~\ref{exi}. In Section~\ref{emqt} we move beyond minimally-coupled matter and analyse various models of QT theories coupled to electromagnetic QT gravities. We show that for some of them there exist universal bounds for curvature invariants---both independent of the mass and the charge of the solutions. We argue that such bounds are lost when the non-minimally coupled matter is turned off and exclusively minimally coupled matter is kept. We conclude with several comments and future directions in Section~\ref{disc}.
Various intermediate calculations regarding the effects of different types of singular matter for several families of qualitatively different QT models are relegated to appendices~\ref{app:blowups_unified} and~\ref{app:fast_decay}. Finally, in Appendix~\ref{app:OS} we remove the staticity condition on the matter models and verify the boundedness of curvature invariants for cosmological spacetimes filled with perfect-fluid matter.

\section{Static Spherically symmetric solutions with matter}\label{sssm}

A general higher-curvature theory built from contractions of the metric and Riemann tensor can be written as  
\begin{equation}
    I=\int \dif^D x \sqrt{|g|} \left[\frac{\mathcal{L}\left (R_{abcd},g^{ef} \right)}{16\pi \GN}+\mathcal{L}_{\rm matter} \right] \,.
\end{equation}
The corresponding equations of motion read~\cite{Padmanabhan:2011ex}
\begin{align}\label{eq:Eab}
    P_a{}^{cde}R_{bcde}-\frac{g_{ab}\mathcal{L}}{2}-2\nabla^c\nabla^d P_{acdb}=8\pi \GN T_{ab} \, , \!
\end{align}
where
\begin{equation}
    P^{abcd}\equiv \left[ \frac{\partial \mathcal{L}}{\partial R_{abcd}}\right]\, ,
\end{equation}
and, in general, the field equations are fourth-order in derivatives.

In what follows we will focus on Quasi-topological (QT) gravities~\cite{Oliva:2010eb,Quasi,Dehghani:2011vu,Ahmed:2017jod,Cisterna:2017umf,Bueno:2019ycr,Bueno:2022res,Moreno:2023rfl,Moreno:2023arp,Bueno:2025qjk}, a broad class of higher-curvature theories with non-trivial representatives at essentially every curvature order in general dimensions $D\geq 5$ (the only exception occurs at $n=D/2$ in even dimensions, for which the corresponding densities are trivial). A key feature of QT gravities is that, despite their higher-derivative nature, their spherically symmetric sector is typically governed by markedly simpler equations and---except for a measure-zero subset---these theories satisfy a Birkhoff theorem, so their most general spherically symmetric vacuum solutions are static and take a Schwarzschild-Tangherlini-like form.

We consider Lagrangians of the form
\begin{equation}\label{genAc}
   \mathcal{L}\left (R_{abcd},g^{ef} \right) = R+\sum_{n=2}^{n_{\rm max}} \widetilde{\alpha}_n \mathcal{Z}_n \, ,
\end{equation}
where $\widetilde{\alpha}_n$ are arbitrary couplings with dimensions of length$^{2(n-1)}$, introducing new length scales in the theory. In this work, we shall work in the limit $n_{\rm max} \to \infty$, for which the vacuum solutions of the corresponding theories are regular black holes under very mild assumptions~\cite{Bueno:2024dgm}. 

The QT densities $\mathcal{Z}_n$ can be obtained at arbitrary order starting from the first five, $\mathcal{Z}_i$, $\{i=1,\dots,5\}$, using the recursive formula~\cite{Bueno:2019ycr}
\begin{align}
\label{recu} \nonumber
\mathcal{Z}_{n+5}=
&\frac{3(n+3)\mathcal{Z}_{1}\mathcal{Z}_{n+4}}{D(D-1)(n+1)}-\frac{3(n+4)\mathcal{Z}_{2}\mathcal{Z}_{n+3}}{D(D-1)n}\\
&+\frac{(n+3)(n+4)\mathcal{Z}_{3}\mathcal{Z}_{n+2}}{D(D-1)n(n+1)}\, .
\end{align}
The explicit form of the first five densities is not particularly illuminating and can be found, for instance, in Eq.~7 of~\cite{Bueno:2024zsx}.

We study general static and spherically symmetric metrics, which can be written as 
\begin{equation}\label{Nf}
    \dif s^2 = - N(r)^2f(r) \dif t^2 + \frac{\dif r^2}{f(r)} + r^2\dif\Omega_{D-2}^2 \,.
\end{equation}
In this setting, we consider quasi-topological gravity minimally coupled to matter. The general form of the stress tensor compatible with spherical and time-reflection symmetry reads
\be \label{eq:Tab}
T_{a}^{b} = {\rm diag} \left(-\rho, p_r, p_t,\dots, p_t\right) \, .
\ee
In a spacetime region where $t$ is the time coordinate, $\rho(r)$ is the mass-energy density, $p_r(r)$ is the radial pressure and $p_t(r)$ is the tangential pressure. The same form of the stress tensor applies also in regions where $f(r) < 0$, such as black hole interiors, though $\rho(r)$ is no longer an energy density in that case.\footnote{Similarly, we will refer to $r$ as the radial coordinate throughout, even inside horizons where it is timelike.} Throughout, we shall refer to $\rho(r)$ as the energy density with the understanding that this name derives from the physical interpretation of $\rho(r)$ in the static coordinate chart. If $p_r = p_t$, then the stress tensor describes a perfect fluid but we shall not in general impose this requirement. Local conservation of the stress tensor implies that
\begin{align} \label{eq_conservationlaw}
    \frac{\dd p_r}{\dd r}
    = &-\frac{1}{2}(\rho+p_r)\frac{\dd \log(N^2f)}{\dd r} 
    \nonumber 
    \\
    &- \frac{(D-2) (p_r - p_t)}{r}\, .
\end{align}
For this system, there are two independent gravitational field equations, which read
\begin{align} \label{eq_QTequations0}
    & \quad \frac{\dif}{\dif r}\left[r^{D-1}h(\psi)\right] = -\frac{16\pi\GN}{(D-2)} r^{D-2}T_{t}^{t} \,, \\ \label{eq_QTequations1}
    & \quad \frac{N'}{rN}\!=\!\frac{8\pi\GN}{(D - 2) f h'(\psi)}\left( T_{r}^{r}-T_t^t \right) \,.
\end{align}
The \textit{characteristic function} $h(\psi)$ encodes all the information about the quasi-topological theory under consideration. It takes the form 
\begin{equation}\label{char_poly}
h(\psi) \equiv \psi + \sum_{n=2}^\nmax \alpha_n \psi^n\, ,  
\end{equation}
where we defined
\begin{equation} \label{eq_f_psi}
    \psi \equiv \frac{1-f(r)}{r^2}\, ,\qquad  \alpha_n \equiv \widetilde{\alpha}_n\frac{(D-2n)}{(D-2)}\, ,
\end{equation}
being $\widetilde{\alpha}_n$ the couplings appearing in the action.

Combined with the generalized Bianchi identity and local conservation of the stress tensor, one finds the angular field equations are satisfied. We therefore have a system of three differential equations for five unknown functions $\{N(r), f(r), \rho(r), p_r(r), p_t(r) \}$. Hence, two more conditions must be provided to obtain a closed system. These can be taken to be the equation of state of the system and a prescription for the pressure anisotropy $p_r-p_t$. If the matter arises from an action principle, then these conditions are automatically provided as a consequence of the matter field equations. An important observation for our work here is that we will be able to characterize the behaviour of the spacetime curvature \textit{without} imposing matter field equations. Hence we shall obtain results that hold independent of the matter model, requiring only certain behaviours for $\rho$ and $p_r$. 

Let us now proceed to obtain the form of $f(r)$ and $N(r)$. It will be convenient to introduce rescaled versions of the energy density and pressure according to
\be 
\{ \varrho, \p_r, \p_t \}  \equiv  \frac{8\pi\GN }{(D-2)} \{\rho,p_r, p_t \}  \, .
\ee
The solution of the first field equation is expressed most simply after the introduction of a new function $S(r)$ defined as~\cite{Hennigar:2020kqt}
\be 
S(r) = \frac{2 \mathsf{M}}{r^{D-1}} + \frac{2}{r^{D-1}} \int_\infty^r x^{D-2} \varrho(x) \dd x \, ,
\ee
where
\be 
\M \equiv \frac{8\pi\GN }{(D-2)\Omega_{D-2}} M
\label{eq:mass}
\ee
with $M$ the ADM mass of the solution. The first field equation then reads
\be 
h(\psi) = S(r)\, ,
\ee
which we can formally solve by introducing the inverse function of $h(x)$ which we denote $H(x)$, \emph{i.e.,} $H(h(x)) = x$,
\be 
\psi = H(S)  \quad \Rightarrow \quad f(r) = 1 - r^2 H(S) \, .
\ee
While the characteristic function $h(x)$ is what is prescribed by the theory, we shall see that it is the inverse function $H(x)$ that is most useful for assessing the regularity of solutions.

For $N(r)$, we have the equation
\begin{align}
\frac{N'}{r N} = \frac{(\p_r + \varrho)}{f h'(\psi)} \, .
\end{align}
Taking the normalization such that $N(\infty) = 1$ and writing things in terms of $H(x)$,\footnote{Here we use the relationship between the derivative of a function and its inverse, $h'(\psi) = 1/H'(S)$.} we can solve the above as
\be \label{eqN}
N(r)\!=\!\exp\!\left[\int_\infty^r \!\frac{x H'(S(x)) \left( \p_r(x) + \varrho(x)\right) }{1 - x^2 H(S(x))} \dd x\right]  \!.
\ee
We note that we have not used the conservation equation in arriving at the above results. Hence they hold generally, independent of the matter equations.

Our main objective is to determine the conditions under which solutions to this system have bounded curvature invariants. To this end, we now introduce the Riemann tensor for the general static and spherically symmetric metric~\eqref{Nf}, which is most conveniently written as
\begin{widetext}
\begin{equation}
 R_{ab}{}^{cd} =\mathcal{R}^{(1)}\tau_{[a}^{[c}  \rho_{b]}^{d]}+\mathcal{R}^{(2)} \tau_{[a}^{[c}  \sigma_{b]}^{d]} +\mathcal{R}^{(3)} \rho_{[a}^{[c} \sigma_{b]}^{d]}+\mathcal{R}^{(4)} \sigma_{[a}^{[c}  \sigma_{b]}^{d]}\,,
\label{eq:RformNf}
\end{equation}
where we have defined
\begin{align}
\mathcal{R}^{(1)}&=-\frac{6 f' N'+2N f''+4f N''}{N}\,, \,\, \mathcal{R}^{(2)}=-\frac{2N f'+4f N'}{r N}\,, \,\, \mathcal{R}^{(3)}=-\frac{2 f'}{r} \,, \,\,\mathcal{R}^{(4)}=2 \frac{(1-f)}{r^2} \,.
\label{eq:RformNfcomp}
\end{align}
Here $\tau_a^b = \delta_a^t \delta^b_t$, $\rho_a^b = \delta_a^r \delta^b_r$, and $\sigma_a^b = \delta_a^b - \rho_a^b - \tau_a^b$ are projectors on the $t$, $r$, and sphere directions, respectively. As they are projectors, we have $\tau_a^c \tau_c^b = \tau_a^b$, $\rho_a^c \rho_c^b = \rho_a^b$, and $\sigma_a^c \sigma_c^b = \sigma_a^b$. Since they project onto orthogonal subspaces we have that the contraction between any two different projectors vanishes, $\tau_a^b \rho_b^c = \tau_a^b \sigma_b^c= \rho_a^b \sigma_b^c=  0$. Finally, the traces satisfy $\tau_a^a = 1 = \rho_a^a$ and $\sigma_a^a = D-2$. 

Any \textit{algebraic} curvature invariant of order $n$---built from contractions of the metric and curvature tensors and products thereof---can be expressed as a homogeneous polynomial of degree $n$ in the $\mathcal{R}^{(i)}$. For instance, the Kretschmann scalar is an $n=2$ invariant and reads
\be 
R_{abcd}R^{abcd} = \frac14\Big[\big(\mathcal{R}^{(1)}\big)^2
+(D-2)\Big(\big(\mathcal{R}^{(2)}\big)^2+\big(\mathcal{R}^{(3)}\big)^2\Big)\Big]
\;+\;
\frac{(D-2)(D-3)}{2}\,\big(\mathcal{R}^{(4)}\big)^2 \, .
\ee
Therefore, a sufficient condition for all polynomial curvature invariants to remain bounded is that each $\mathcal{R}^{(i)}$ be bounded. This is in fact also necessary: the $\mathcal{R}^{(i)}$ are real, and there exist invariants (\emph{e.g.,} the Kretschmann scalar above) that involve only even powers of the $\mathcal{R}^{(i)}$ with strictly positive numerical coefficients, precluding any generic cancellations among terms.

We must evaluate the above curvature components on the metrics under consideration. Noting the identities
\begin{align}
    S' &= - \frac{(D-1) S}{r} + \frac{2 \varrho}{r} \, ,
    \qquad
    S'' = \frac{D(D-1) S}{r^2} - \frac{2 D \varrho}{r^2} + \frac{2 \varrho'}{r} \, ,
\end{align}
and using the solution for $N(r)$ and $f(r)$ presented above, we can obtain the following expressions for the various Riemann tensor components:
\begin{align}\label{eqn:RiemComps}
    \mathcal{R}^{(1)} &= 4 H(S) + 2 (D-1)(D-4) S H'(S) - 4 (D-4) \varrho H'(S) + 4 r \varrho' H'(S) + 2 ((D-1)S - 2 \varrho)^2 H''(S)
    \nonumber\\ 
    &- 6 r \big[((D-1)S - 2 \varrho) H'(S) - 2 H(S)\big] W - 4 (1 - r^2 H(S))(W' + W^2) \, ,
    \\
    \mathcal{R}^{(2)} &= 4 H(S) - 2 (D-1) S H'(S) -4 \p_r H'(S) \, ,
    \\
    \mathcal{R}^{(3)} &= 4 H(S) - 2 (D-1) S H'(S) + 4 \varrho H'(S) \, ,
    \\
    \mathcal{R}^{(4)} &= 2 H(S) \, ,
\end{align}
where we introduced the short hand
\be 
W(r) = \frac{N'(r)}{N(r)} = \frac{r  H'(S) (\p_r + \varrho)}{1 - r^2 H(S)} \, .
\ee
The $\mathcal{R}^{(i)}$ are complicated. To obtain a better understanding of the boundedness conditions, we shall study different scenarios, beginning with the vacuum case. 

To help orient the discussion, let us introduce two representative classes of characteristic functions. The Hayward-like family of resummations is given by
\be \label{eqn:hayward_model}
h(x) = \frac{x}{\left(1- \alpha^{\rm N} x^{\rm N} \right)^{1/{\rm N}}} \quad \Rightarrow \quad H(x) = \frac{x}{\left(1+ \alpha^{\rm N} x^{\rm N} \right)^{1/{\rm N}}}
\ee
where ${\rm N}$ is a positive integer and $\alpha$ is the length-squared scale of the theory. In the case ${\rm N} = 1$, the vacuum solution of the theory is the $D$-dimensional Hayward regular black hole. A second example is the $\arctanh$ resummation for which we have
\be \label{eqn:arctanh_model}
h(x) = \frac{\arctanh(\alpha x)}{\alpha} \quad \Rightarrow \quad   H(x) = \frac{1}{\alpha}\tanh(\alpha x) \, .
\ee
As it will be important later, let us note that as $|x| \to \infty$ in the Hayward-like models we have the inverse function decays polynomially as $H'(x) \sim |x|^{-\beta}$ and $H''(x) \sim |x|^{-(\beta+1)}$ with $\beta = {\rm N} + 1$,\footnote{Unless otherwise stated, $f(x)\sim g(x)$ denotes asymptotic equivalence up to multiplicative constants (not necessarily with unit ratio). Whenever the multiplicative constant is relevant, it will be written explicitly.} while in the $\arctanh$ model the decay is super-polynomial with $H'(x), H''(x) \sim \exp{(-2 |x|)}$ as $|x| \to \infty$. There are, of course, many other possible models---see {\it e.g.}~\cite{Bueno:2024dgm, Hennigar:2020kqt} for additional examples. However, here the main point is to illustrate that models with regular black holes exist for which the inverse function decays either polynomially or super-polynomially for large argument.

\end{widetext}

\section{Limiting Curvature Hypothesis in Vacuum}\label{lchv}

In vacuum we have $\varrho = \p_r = \p_t = 0$. The field equations simplify so that we have $N(r) = 1$ and 
\be 
f(r) = 1 - r^2 H(S) \quad \text{with}\quad S(r) = \frac{2 \mathsf{M}}{r^{D-1}} \, .
\ee 
The $\mathcal{R}^{(i)}$ also simplify to
\begin{align}
    \mathcal{R}^{(1)} &= 4 H(S) + 2 (D-1)(D-4) S H'(S) 
    \nonumber
    \\ \label{eq:R1f} &+2 (D-1)^2 S^2 H''(S) \, ,
    \\
    \label{eq:R2f}
    \mathcal{R}^{(2)} &= \mathcal{R}^{(3)} = 4 H(S) - 2 (D-
    1) S H'(S)  \, ,
    \\
    \label{eq:R3f}
    \mathcal{R}^{(4)} &= 2 H(S) \, .
\end{align}
The crucial observation here (which was first made in~\cite{Frolov:2024hhe} in a different notation) is that in vacuum the $\mathcal{R}^{(i)}$ involve only $H(S)$, its derivatives, and $S$---other parameters of the solution (such as the mass) make no explicit appearance. Since any algebraic curvature invariant will be a homogeneous function of the $\mathcal{R}^{(i)}$, any such invariant will inherit this property. As a result, if any polynomial curvature invariant is bounded, then that bound cannot depend on the mass in vacuum. This is the essence of Markov's limiting curvature hypothesis: a universal upper bound on the curvature. 

So, then, under which circumstances are the algebraic curvature invariants bounded? As we have described, an algebraic curvature invariant of order $n$ will necessarily be a homogeneous polynomial of order $n$ in $\mathcal{R}^{(i)}$. In vacuum, the $\mathcal{R}^{(i)}$ themselves are bounded provided that each of 
\be \label{vacuum_inv}
\{H(S)\,, \,\, S H'(S) \, , \,\, S^2 H''(S) \}
\ee
are all bounded on their corresponding domain. In vacuum, $S \in (0, \infty)$ and this gives the domain of $H$. The inverse function $H$ must be such that it is bounded on that domain. Combined with the boundedness of $H(x)$, the second and third conditions tell us that $H'(x)$ and $H''(x)$ are not permitted to have poles of order 1 and 2, respectively, at $x = 0$. More importantly, those conditions also tell us that as $x \to \infty$ we must have $|H'(x)|$ decay faster than $1/|x|$ and $|H''(x)|$ decay faster than $1/x^2$. If these conditions are met, then all polynomial curvature invariants will be bounded (and therefore regular).\footnote{The marginal decay $H'(x)\sim 1/|x|$ would keep $xH'(x)$ bounded but forces $H(x)\sim \log|x|$, contradicting bounded $H(x)$ (and yielding divergent curvature).}  The example resummations we gave in Eq.~\eqref{eqn:hayward_model} and~\eqref{eqn:arctanh_model} satisfy these conditions. 

The regularity conditions are most easily expressed in terms of the inverse function $H(x)$, but what the theories actually give us is the characteristic function $h(x)$. So, what do these conditions imply for $h(x)$? The existence and uniqueness of the inverse function $H(x)$ can be ensured provided that $h(x)$ is strictly monotonic on its domain. The existence of an Einstein gravity limit with positive Newton constant further requires $h'(0) = 1$, and hence $h(x)$ should be monotone \textit{increasing}. The requirement that  $H(x)$ has a range which is bounded implies that the domain of $h(x)$ is finite. Moreover, the fact that the (vacuum) domain of $H(x)$ is $[0, \infty)$ indicates that $h(x)$ must have this interval as its range, on a finite domain. Hence, the function $h(x)$ should be a strictly monotone increasing function defined on an interval $x \in [0, b)$ with a divergence at the endpoint of the interval. The conditions on the decay of the derivatives of $H(x)$ can be translated into the following endpoint requirements on $h(x)$ and its derivatives:
\begin{align}
    \lim_{x \to b^-}\left|\frac{h(x)}{h'(x)}\right| \!= 0   \, , \!\quad 
    \quad \lim_{x \to b^-}\left|\frac{h(x)^2\,h''(x)}{\big(h'(x)\big)^3}\right|\!= 0 .
\end{align}
It is difficult to translate these requirements into useful necessary/sufficient conditions on the coupling constants themselves and we shall not do that here.\footnote{A set of such conditions were proposed in~\cite{Bueno:2024dgm}, but those conditions were incorrect. A correct set of \textit{sufficient} conditions on the coupling constants appear in~\cite{Fernandes:2025fnz}. We thank Pedro G.~S.~Fernandes for bringing this to our attention. None of the results of~\cite{Bueno:2024dgm} or follow-ups are affected by this correction.}

\section{Matter satisfying \texorpdfstring{$T_t^t = T_r^r$}{T\_t\^t=T\_r\^r}}\label{mstt}

Before addressing the most general case, let us first consider minimally coupled matter satisfying $T_t^{t}=T_r^{r}$. This condition implies $\p_r=-\vrho$, and then~\eqref{eqN} yields $N'(r)=0$, so $N(r)$ is a constant that we may set to unity. Hence, for $T_t^{t}=T_r^{r}$ the most general static solution is characterized by a single metric function, $f(r)$. This constraint may seem severe but, for example, Maxwell theory as well as any theory of nonlinear electrodynamics satisfies this property~\cite{Hennigar:2020kqt}.

Under these assumptions, the $\mathcal{R}^{(i)}$ simplify considerably and read:
\begin{align}
    \mathcal{R}^{(1)} &= 4 H(S) + 2 (D-1)(D-4) S H'(S) 
    \nonumber 
    \\
    &- 4 (D-4) \varrho H'(S) + 4 r \varrho' H'(S) 
    \nonumber 
    \\
    &+ 2 ((D-1)S - 2 \varrho)^2 H''(S)  \, ,
    \\
    \mathcal{R}^{(2)} &  = \mathcal{R}^{(3)} = 4 H(S) - 2 (D-1) S H'(S) 
    \nonumber 
    \\ 
    &+ 4 \varrho H'(S) \, ,
    \\
    \mathcal{R}^{(4)} &= 2 H(S) \, .
\end{align}
Once minimally coupled matter is introduced, it is already clear that any parameter-independent curvature bounds will generally be lost. Indeed, the curvature coefficients now acquire explicit dependence on $r$ and on the matter profile $\varrho$, in addition to $H$ and $S$. This raises a more basic question: do the curvature invariants remain bounded at all? If so, what (if any) conditions must the matter sector satisfy in order to ensure bounded curvature?

A new feature arises in the presence of matter. Just as in vacuum, the range of $h(x)$ (and therefore the domain of $H$) is determined by $S(r)$. In vacuum $S(r) \in (0, \infty)$ but with matter this domain can differ as $S(r)$ can take on negative values. In fact, not only \textit{can} this happen, but if the matter energy density is \textit{positive} this will happen generally. Therefore, let us make a few assumptions to allow progress. First, let us suppose that the domain of $H(x)$ is the real numbers. This means we are allowing for $S(r) \in \mathbb{R}$. In certain cases this assumption will be stronger than necessary but this will allow us to avoid worrying here about the range of $S(r)$ for matter models in a case-by-case analysis.  Our assumption that the domain of $H(x)$ is the real numbers therefore ensures that the original function $h(x)$ was invertible for all $S(r) \in \mathbb{R}$. This is not true for the Hayward(-like) theories in Eq.~\eqref{eqn:hayward_model} with \textit{odd} ${\rm N}$, but it does hold, for example, for 
\be \label{N2H}
h(x) = \frac{x}{\sqrt{1-\alpha^2 x^2}} \, \, \Rightarrow \,\, H(x) = \frac{x}{\sqrt{1+ \alpha^2 x^2}}
\ee
as well as for many others~\cite{Bueno:2024dgm, Hennigar:2020kqt} including the Hayward(-like) theories in Eq.~\eqref{eqn:hayward_model} with \textit{even} ${\rm N}$ as well as the $\arctanh$ model~\eqref{eqn:arctanh_model}. Let us further suppose that we have chosen the theory such that the polynomial curvature invariants are all bounded in vacuum, satisfying the limiting curvature hypothesis in that case. Finally, we assume that the matter is well-behaved in the sense that $\vrho$ decays sufficiently fast near infinity that the integral required for $S(r)$ converges there. 

We then find  that boundedness of $\mathcal{R}^{(i)}$ requires boundedness of
\begin{align} \label{eqn:matter_bounds1}
\nonumber
&\vrho(r) H'(S) \,,  &r \vrho'(r) H'(S) \, ,
\\
&\vrho(r)^2 H''(S) \, , &\vrho(r) S H''(S)\,.
\end{align}
The question is whether the vacuum boundedness conditions imply these relationships, or whether the boundedness of these quantities requires additional restrictions on the matter content of the theory. What makes this somewhat nontrivial is that $S(r)$ depends itself on $\vrho$ in a complicated manner and this can lead to surprising consequences. We collect our calculations in Appendices~\ref{app:blowups_unified} and~\ref{app:fast_decay} and comment here on the results. 

A little thought reveals that the vacuum boundedness conditions on their own \textit{do not in general} imply boundedness of the quantities in~\eqref{eqn:matter_bounds1}. The underlying reason is simple: the resummation suppresses matter contributions only when the argument $S(r)$ is driven to large magnitude, $|S|\to\infty$, so that $H'(S)$ and $H''(S)$ are small. If instead the density diverges while $S(r)$ remains finite, then $H'(S)$ and $H''(S)$ evaluate to ordinary finite numbers and cannot compensate the blow-up of $\vrho$ or $\vrho'$.

A simple counterexample is to take $\vrho(r)$ to have an \textit{integrable} singularity at some $r=r_p>0$. In such a case, $S(r_p)$ remains perfectly finite yet the density is divergent and hence the curvature terms grow unbounded, reflecting the divergent matter energy density. To make this more concrete, let us take an explicit example. Suppose we take the ${\rm N}=2$ Hayward model, determined by (\ref{N2H}),
which meets all the conditions we have specified. For the density profile, let us take
\be
\vrho(r)=\frac{\M_p}{\pi r^{D-1}}\sqrt{\frac{r_p}{|r-r_p|}} \, ,
\ee
where $\M_p$ is some constant with units of mass. Then, it is easy to check that
$S(r_p)=(2\mathsf{M}-2\M_p)/r_p^{D-1}$, which is clearly finite. Moreover,
\be
H'(x)=\frac{1}{(1+\alpha^2 x^2)^{3/2}}
\ee
which is also finite when evaluated for $x=S(r_p)$. Hence, $\vrho(r)\,H'(S)$ diverges at $r=r_p$---and it is a simple exercise to check that $\vrho^2 H''(S)$, $\vrho\,S\,H''(S)$ and $r\vrho'(r)\,H'(S)$ are also divergent in this example---and therefore curvature scalars will generically be divergent at $r=r_p$ in such a model.

We have concluded that even in the simplest setting of minimally coupled matter satisfying $T^t_t=T^r_r$, the boundedness of the vacuum curvature invariants does not imply the boundedness of those same invariants in the presence of matter. However, this is not particularly surprising: at this point we have demanded essentially nothing from the matter sector. One would like to understand what additional constraints must be placed on $\vrho(r)$ so that curvature invariants \emph{do} remain bounded.

Clearly, if the matter sector itself satisfies certain boundedness properties, \emph{e.g.}, $|\vrho|<\vrho_{\rm max}$ along with the boundedness of $\vrho'(r)$, then the curvature invariants will also be bounded. However, the full situation is actually (and perhaps surprisingly) far less constraining than this. The key point is whether the matter blow-up forces $|S|\to\infty$, since only then does the large-$|S|$ decay of $H'(S)$ and $H''(S)$ become effective.

First, if $\vrho(r)$ is finite for all $r>0$ and exhibits a power-law divergence at the center,
\be
\vrho(r)\sim \frac{1}{r^{\gamma_1}} \qquad (r\to 0^+;\ \gamma_1>0)\, ,
\ee
then $|S(r)|\to\infty$ as $r\to 0$ (either because the vacuum term $2\mathsf{M}/r^{D-1}$ dominates or because the matter contribution does). In this regime the resummation suppresses the potentially dangerous combinations in~\eqref{eqn:matter_bounds1}. In particular, assuming the same large-$|S|$ behaviour that is sufficient to ensure bounded curvature in vacuum (namely $H'(S)\sim |S|^{-\beta}$ and $H''(S)\sim |S|^{-(\beta+1)}$ with $\beta>1$ or faster), one finds that all density-driven terms in~\eqref{eqn:matter_bounds1} remain bounded as $r\to 0^+$ when $T^t_t=T^r_r$. We provide the detailed estimates in Appendix~\ref{app:blowups_unified}.

Second, one can even consider more severe circumstances in which $\vrho(r)$ has a \textit{non-integrable} power-law singularity at some finite radius $r=r_p>0$,
\be
\vrho(r)\sim \frac{1}{(r-r_p)^{\gamma_1}} \qquad (r\to r_p^+;\ \gamma_1>1)\,.
\ee
In this case the integral defining $S(r)$ diverges, so $|S|\to\infty$ as $r\to r_p$, and the resummation can again suppress the matter blow-up. For resummations with polynomial large-$|S|$ decay (as in the Hayward-like models), Appendix~\ref{app:blowups_unified} shows that bounded curvature is ensured provided that $H'(S)\sim |S|^{-\beta}$ with $\beta>1$ and that the blow-up is sufficiently strong:
\be\label{colo}
\gamma_1 \ge \frac{\beta+1}{\beta-1}\, .
\ee
The direction of this inequality is not a typo: for fixed $\beta$ there is a \textit{lower} bound on the strength of the matter singularity required for it to be regularised. In cases where derivatives of the inverse function decay faster than polynomial for large argument, the above becomes a sufficient but not necessary condition: polynomial matter blow-ups are regularised for any $\gamma_1 > 1$ in such a case, as explained in Appendix~\ref{app:fast_decay}. Schematically, these cases work because the divergences of $\vrho$ occur simultaneously with divergences of $S(r)$, so that, for example, $\vrho\,H'(S)$ inherits the same asymptotic behaviour as $S\,H'(S)$, and similarly for the other terms in~\eqref{eqn:matter_bounds1}.

Finally, the marginal finite-radius case $\gamma_1=1$ is more subtle: $|S|$ diverges only logarithmically, and the outcome depends on the \textit{rate} of decay of $H'(S)$ and $H''(S)$ (see Appendix~\ref{app:fast_decay} for a summary). For polynomial decays, the suppression is not strong enough and the curvature invariants diverge, as shown in Appendix~\ref{app:blowups_unified}.

Putting these cases together, we find a simple organizing principle in the $T^t_t=T^r_r$ sector: singular matter leads to singular curvature precisely when the singularity is mild enough that it does not drive $|S|\to\infty$ (as for integrable finite-radius blow-ups). Conversely, when the matter or vacuum blow-up forces $|S|$ large (as at the center, or for non-integrable finite-radius divergences), the resummation can suppress the would-be divergences and yield bounded curvature. Thus we reach the somewhat surprising conclusion that quasi-topological gravity gives rise to regular geometries sourced by singular matter provided that the matter energy density is \textit{sufficiently singular}.

\section{General Case}\label{gc}

Let us now move on to the general case, for which we will not have $\p_r(r) =- \varrho(r)$. We then need to deal with the general components of the Riemann tensor given in~\eqref{eqn:RiemComps}. In addition to the boundedness of the vacuum curvature invariants, we must also have the boundedness of the terms in~\eqref{eqn:matter_bounds1} along with the new terms
\begin{align}\label{eqn:matter_bounds2}
  \p_r(r) H'(S) &\, , &  rW(r) H(S) &\, , & rW(r)\varrho H'(S) \, , \nonumber \\
  rW(r) S H'(S) &\, , & f(r) W'(r) &\, , & f(r) W(r)^2 \, .
\end{align}

Before turning to the regimes where the resummation can suppress matter divergences, it is useful to isolate a few situations in which curvature singularities are essentially unavoidable. First, if $\varrho$ has an \emph{integrable} singularity at some $r_0>0$, then $S(r_0)$ remains finite and $H'(S(r_0))$ is just a finite number, so terms such as $\varrho\,H'(S)$ will diverge and curvature invariants will be singular. Similarly, if $\p_r$ has a singularity at some $r_0\geq0$ while $\varrho(r_0)$ remains either finite or has an integrable singularity, then terms such as $\p_r\,H'(S)$ diverge and the geometry is singular. This latter case includes, for example, constant-density stars with divergent internal pressure~\cite{Bueno:2025tli}.

So far, these are singularities driven directly by corresponding (mild) divergences in the matter sector. Let us now assume that the energy density and radial pressure are singular at $r=0$ with behaviour
\be
\varrho \sim \frac{1}{r^{\gamma_1}} \, ,\qquad
\p_r \sim \frac{1}{r^{\gamma_2}}\, ,
\ee
with $\gamma_1>0$ and $\gamma_2\in\mathbb{R}$. As we discovered in the previous section (see Appendix~\ref{app:blowups_unified} for details), all the terms involving the energy density in~\eqref{eqn:matter_bounds1} are automatically finite. The only subtleties arise from the pressure-dependent terms, and as we show in Appendix~\ref{app:blowups_unified}, the solutions will be regular in the case of polynomial decay provided that
\be \label{eq:genconditions}
\begin{cases}
\gamma_2 \le \beta (D-1) \quad &\text{if} \quad \gamma_1 \le D-1 \, ,
\\[2pt]
\gamma_2 \le \beta \gamma_1 \quad &\text{if} \quad \gamma_1 > D-1 \, .
\end{cases}
\ee

Because $\beta>1$ by the vacuum boundedness conditions, this means that (within these sufficient conditions) the pressure may be either less or more singular than the energy density at $r=0$ while maintaining a regular geometry. On the other hand, if the inverse function decays faster than polynomial, the curvature invariants will be bounded with no further constraints on $\gamma_2$ (see Appendix~\ref{app:fast_decay}). 

Let us now suppose that both $\varrho$ and $\p_r$ have power-law singularities at some $r=r_p>0$, characterized by different strengths,
\be \label{eq:behav}
\varrho \sim \frac{1}{(r-r_p)^{\gamma_1}}\, ,\
\p_r \sim \frac{1}{(r-r_p)^{\gamma_2}} \quad (r\to r_p^+)\, ,
\ee
with $\gamma_1 > 1$, $\gamma_2\in\mathbb{R}$ and $f(r_p)\neq 0$. We know from the above that all terms in the $\mathcal{R}^{(i)}$ involving factors of $\varrho$ will be bounded provided that 
\be\label{eqn:gam_boundv1}
\gamma_1\ge \frac{\beta+1}{\beta-1}\, .
\ee
Examining the additional pressure- and $W$-dependent terms, one finds (Appendix~\ref{app:blowups_unified}) that if $\gamma_2\le \gamma_1$ then all of the terms are bounded without any further constraint on $\gamma_2$. However, if $\gamma_2>\gamma_1$ then one must also have
\be \label{eq:genblowup}
\gamma_2 \leq \beta(\gamma_1-1)-1
\ee
to ensure boundedness of all terms. 
These conditions are derived in Appendix~\ref{app:blowups_unified}. In the case of faster than polynomial decay, then there are no constraints imposed on $\gamma_2$ and the configurations will be regular in general. 

The only remaining case to examine is the marginal one, corresponding to $\gamma_1=1$ in~\eqref{eq:behav}. In that case, the behaviour of $S(r)$ as we approach $r_p$ reads
\begin{equation}
    S(r)\sim \lambda_{\rm log} \log(r-r_p) \quad (r\to r_p^+) \,,
\end{equation}
where $\lambda_{\rm log} = \frac{2}{r_p} \lim_{r\to r_p^+}(r-r_p)\varrho(r)$. This behaviour causes the matter fields to diverge exponentially in $|S|$, resulting in a curvature singularity, since no finite $\beta$ can suppress this growth. See Appendix~\ref{app:blowups_unified} for more details. However, as we discuss in Appendix~\ref{app:fast_decay}, if the decay of the inverse function is super-polynomial, in many circumstances this marginal blow-up can also be regularised. 

A qualitatively different divergence can arise in the general case due to the $W$-dependent terms in $\mathcal{R}^{(1)}$. In particular, the contribution
\be
\frac{\p_r+\varrho}{f(r)}
\ee
is dangerous. In the presence of a horizon $r=r_h$ where $f(r_h)=0$, one generically finds divergent curvature invariants in any theory of gravity unless the \emph{horizon regularity condition}
\be
\p_r(r_h)=-\varrho(r_h)
\ee
holds (assuming $\rho$ and $p_r$ are themselves finite there). Physically, this reflects a ``blueshift'' divergence. To see this, consider the local energy density recorded by a freely falling radial observer. The $D$-velocity satisfies $u_a u^a=-1$ with nontrivial components
\be
u^t=\frac{E}{N^2 f}\, ,\qquad
u^r=-\sqrt{\frac{E^2}{N^2}-f}\, ,
\ee
and the local energy density perceived by this observer is
\be
\rho_{\rm loc}=T_{ab}u^a u^b=\frac{E^2(p_r+\rho)}{N^2 f}-p_r\, ,
\ee
which diverges at $f=0$ unless $p_r=-\rho$ at the horizon. In most constructions of static black holes sourced by matter, $p_r(r_h)=-\rho(r_h)$ is imposed as a regularity condition~\cite{Visser:1992qh, Padmanabhan:2002sha,Medved:2004ih}.

Let us now consider an alternative possibility, motivated by the previous results: \emph{sufficiently singular} matter can sometimes yield bounded curvature invariants in quasi-topological gravity. Specifically, we can ask whether $\varrho$ and $\p_r$ may blow up \emph{at} a horizon while the geometry nevertheless remains regular. Mathematically, one finds that this is only possible provided that (i) the horizon is an inner/cosmological horizon and (ii) $\varrho$ is negative in its vicinity. The reason is that for matter to diverge at a horizon in a manner that can be suppressed in the geometry, one must have $S(r)\to +\infty$ as $r\to r_h$, which requires a locally negative $\varrho$, and the resulting near-horizon form of $f(r)$ forces $f'(r_h)<0$.

To see this explicitly, suppose that near $r=r_h$ the matter behaves as
\be
\varrho(r)\sim -\frac{A}{(r-r_h)^{\gamma_1}}\, ,\quad
\p_r(r)\sim \frac{B}{(r-r_h)^{\gamma_2}}\, ,
\ee
with constants $A,B$ and exponents $\gamma_1,\gamma_2$ such that $\gamma_1>1$. (Note that the marginal $\gamma_1=1$ results in a curvature singularity for polynomially decaying $H(x)$, as in the finite-radius case.) Extracting the most divergent part of $S(r)$ near $r=r_h$ we find
\be
S(r)\sim \frac{2A}{(\gamma_1-1)r_h}\,\frac{1}{(r-r_h)^{\gamma_1-1}}\, .
\ee
Clearly, if $A>0$ then $S(r)\to +\infty$ as $r\to r_h$, while $S(r)\to -\infty$ if $A<0$. Let us denote
\be
\psi_\pm=\lim_{x\to \pm\infty}H(x)\, .
\ee
Then
\be
f(r_h)=1-r_h^2\psi_\pm\, .
\ee
Requiring $f(r_h)=0$ implies $\psi_\pm=1/r_h^2$, which is only consistent when $\psi_\pm>0$. For the relevant examples under consideration ({\it i.e.}, monotonically increasing and bounded functions $H(S)$ for every $S \in \mathbb{R}$, like those in~\eqref{eqn:hayward_model} and~\eqref{eqn:arctanh_model}) this selects $\psi_+$, and therefore we must have $S(r)\to +\infty$ as $r\to r_h$, which is where the requirement $A>0$ comes from.\footnote{Note that for an inner horizon residing in a black hole interior or ``outside'' a cosmological horizon, $t$ is not a time coordinate and hence $\rho(r)$ is not the matter energy density. The requirement of $A > 0$ remains perfectly consistent with positive local energy density.}

What does this imply for the boundedness of curvature invariants? We can leverage without modification the results of the previous section to conclude that all terms appearing in~\eqref{eqn:matter_bounds1} are bounded provided that
\be\label{eqn:gam_bound}
\gamma_1\ge \frac{\beta+1}{\beta-1}\, ,
\ee
for polynomially decaying inverse functions, while $\gamma_1 > 1$ suffices if the resummation decays super-polynomially.  In any event, we note \emph{en passant} that~\eqref{eqn:gam_bound} forces
\be
f(r)\approx -\frac{2(r-r_h)}{r_h}+\cdots\, ,
\ee
so the horizon must be an inner/cosmological one.

On the other hand, one finds
\begin{align}
f(r)W(r)^2&\sim \frac{C_1}{(r-r_h)^{2 \theta_1+1}}+\frac{C_2}{(r-r_h)^{2 \theta_2+1}}\, , \\
f(r)W'(r)&\sim \frac{C_3}{(r-r_h)^{\theta_1+1}}+\frac{C_4}{(r-r_h)^{\theta_2+1}}\, ,
\end{align}
where $\theta_1=\gamma_1-\beta(\gamma_1-1)$, $\theta_2=\gamma_2-\beta(\gamma_1-1)$ and $C_1$, $C_2$, $C_3$ and $C_4$ are constants depending on $A$, $B$ and other factors. If $\gamma_2 \leq \gamma_1$,~\eqref{eqn:gam_bound} is enough to ensure the boundedness of $f W^2$ and $f W'$. If $\gamma_2>\gamma_1$, $W$ is bounded provided~\eqref{eqn:gam_bound} holds and
\begin{equation}
    \gamma_2\leq \beta(\gamma_1-1)-1\,.
    \label{eq:condgammarh}
\end{equation}
We therefore conclude that, despite a slightly more involved analysis for the case of a blow-up at a horizon, the boundedness conditions are exactly the same as in the more general finite-radius blow-up case (see~\eqref{eqn:gam_boundv1} and~\eqref{eq:genblowup}).

It is then relatively easy to verify that all terms entering the $\mathcal{R}^{(i)}$ are individually bounded and hence the solution is regular at $r=r_h$. We therefore see another example of how quasi-topological gravity can yield regular geometries in cases where the matter fields are \textit{sufficiently} singular: the blueshift singularity described above generically results in a singular horizon \textit{except} in the case where the matter fields are themselves singular at the horizon. This curious result may have implications for understanding the mass inflation instability of regular black holes in quasi-topological gravity---see \cite{Frolov:2026rcm}. 

Above we have assumed that $N(r_h)$ takes on a finite constant value at a zero of $f(r_h)$. One may wonder what happens if this is no longer the case. For example, if $N(r)$ exhibits a $N(r)\sim 1/\sqrt{r-r_h}$ blow-up at $r_h$, then the interpretation is different and represents a breakdown of the chart~\eqref{Nf}. In such a case, it is easy to verify that the metric is not singular and the surface $r = r_h$ represents a kind of `throat'. On the other hand, if $N(r)\sim (r-r_h)^\sigma$ for any $\sigma \neq 0, -\frac{1}{2}$, one sees that the geometry will be singular at $r=r_h$. 

The discussion pertaining to horizons has so far focused on the case where the inverse function exhibits polynomial decay at large argument. When the inverse function decays in super-polynomial fashion, the constraints for horizons are identical to those for matter singularities at finite radius---see Appendix~\ref{app:fast_decay}.  In particular, sufficiently strong super-polynomial decay can regularise even the case of a marginal ($\gamma_1 = 1$) matter divergence at an inner horizon, depending on the details of the matter model.

\section{Examples}\label{exi}

We now illustrate the general results obtained so far with explicit examples. We consider electrodynamics and a minimally coupled scalar field. 

\subsection{Electrodynamics}

Let us restrict ourselves to purely electric spherically symmetric ans\"atze. In this case, a general second-order theory of electrodynamics $\mathcal{L}^{\rm EM}(s)$ can be written as a general function of the electromagnetic invariant 
\be 
s = - \frac{ 1}{4} F_{ab} F^{ab} \, .
\ee
The corresponding stress tensor of such a theory takes the form of~\eqref{eq:Tab} with
\begin{align}
\rho &= 2 s \mathcal{L}^{\rm EM}_s - \mathcal{L}^{\rm EM} \, , 
\\ 
p_r &= \mathcal{L}^{\rm EM} - 2 s \mathcal{L}^{\rm EM}_s 
\, , 
\\
p_t &= \mathcal{L}^{\rm EM} \, ,
\end{align}
where $\mathcal{L}^{\rm EM}_s = \partial_s \mathcal{L}^{\rm EM}$. Hence, any theory of electrodynamics satisfies the $T_t^t = T_r^r$ condition in the entire spacetime. 

The simplest and most interesting example is Maxwell electrodynamics, for which $\mathcal{L}^{\rm EM}(s) = s$. The Maxwell equations are solved by $F = \dd A$ with
\be 
A = \sqrt{\frac{(D-2)}{16\pi G_{\rm N}   (D-3)}} \frac{Q}{r^{D-3}} \dd t \, ,
\ee
where $Q$ is a constant of integration. We have
\be 
\varrho = \frac{D-3}{2} \frac{Q^2}{r^{2(D-2)}} = - \p_r
\ee
and hence
\be 
S(r) = \frac{2 \mathsf{M}}{r^{D-1}} - \frac{Q^2}{r^{2(D-2)}} \, . 
\label{eq:selec}
\ee
For fixed $(\mathsf{M},Q)$, the function $S(r)$ takes values in $(-\infty, S_{\rm max}]$ with 
\be 
S_{\rm max} = \frac{D-3}{D-2}\,\mathsf{M}\left(\frac{(D-1)\mathsf{M}}{(D-2)Q^2}\right)^{\frac{D-1}{D-3}} \, .
\ee 
Since $\mathsf{M}$ and $Q^2$ are freely specifiable, positive real numbers, the union of these ranges over all allowed parameter choices is $\mathbb{R}$. Hence, when quasi-topological gravity is coupled to Maxwell electrodynamics, regularity requires resummations for which $H(x)$ has $\mathbb{R}$ as the domain. This holds true for the Hayward-like models with ${\rm N}$ even and the $\arctanh$ model we have considered here. 

The matter fields have power-law divergent energy density and pressure at $r = 0$ with the same exponents $\gamma_1 = \gamma_2= 2(D-2)$. Hence by our general results the regularity of the solution is ensured by the vacuum regularity conditions, provided $H(S)$ has the domain described above. We refer to~\cite{Hennigar:2020kqt} for additional details and examples of regular black holes in quasi-topological gravity coupled to nonlinear electrodynamics.

In another vein, we can actually argue that regular black hole solutions in these theories will not satisfy the limiting curvature hypothesis. To this end, let us write $S_0=\frac{2 \mathsf{M}}{r^{D-1}}$ and $q^2 S_0^\sigma=\frac{Q^2}{r^{2(D-2)}}$ with $\sigma=2 \frac{(D-2)}{D-1}$ and $q^2 (2\mathsf{M})^\sigma=Q^2$ (we consider $\mathsf{M}>0$). Writing $H(S)=H_q(S_0)$, it is clear from~\eqref{eq:R1f},~\eqref{eq:R2f} and~\eqref{eq:R3f} that all curvature invariants will be a function of $\{H(S_0), S_0 H_q'(S_0),S_0^2 H_q''(S_0)\}$, where prime denotes derivative with respect to $S_0$. Let $S_0^{(0)}=q^{-2/(\sigma-1)}$ be the value of $S_0$ such that $S=0$. Direct computation shows that
\begin{equation}
H_q'\left (S_0^{(0)} \right)=H'(0)(1-\sigma)\,.
\end{equation}
Now, as $2 \geq \sigma \geq 4/3$ if $D \geq 4$, we have that $S_0^{(0)} \rightarrow +\infty$ as $q^2 \rightarrow 0$, while $H'(0) (1-\sigma)$ needs not to be small. As a result, $S_0^{(0)} H_q'\left (S_0^{(0)} \right )$ can become arbitrarily large as $q^2 \rightarrow 0$, implying that Markov's proposal will not be satisfied. Although the failure of the limiting curvature hypothesis arises as one considers zero-charge configuration, note that neutral regular black holes will, of course, satisfy Markov's proposal. These ideas can be reconciled by noting that $S \neq 0$ for vanishing charge, so that no $S_0^{(0)}$ exists.

\subsection{Scalar field}

The action of a scalar field $\phi$ minimally coupled to gravity reads
\be 
\mathcal{L}^{\rm field} = -\frac{1}{2} \partial_a \phi \partial^a \phi - V(\phi) 
\ee
where $V(\phi)$ is the scalar potential. In spherical symmetry, the stress tensor that follows from this action takes the form of~\eqref{eq:Tab} with
\begin{align} 
\rho &= \frac{1}{2} f(r) (\phi')^2 +  V \, , \quad  p_r = \frac{1}{2} f(r) (\phi')^2 - V \, , 
\nonumber \\
p_t &= -\left(\frac{1}{2} f(r) (\phi')^2 + V \right) \, .
\end{align}
In this case we have $T_t^t \neq T_r^r$ and in general the solutions will have $N(r) \neq 1$. Because of the form of the stress tensor, provided that $f(r) (\phi')^2 \to 0$ at any horizon we will have $p_r(r_h) = - \rho(r_h)$ and the horizon regularity condition will be satisfied. On the other hand, it could happen that $\phi'(r_h)$ blows up, in which case horizon regularity would depend on the strength of the blow-up.

To understand this better, let us examine the behaviour of the scalar in the vicinity of a horizon (which may be an event or inner horizon). The scalar equation of motion reads
\be 
\frac{1}{N r^{D-2}} \left(N r^{D-2} f \phi'\right)' = V_{,\phi}(\phi) \, ,
\ee
where a ``$\,'\,$'' denotes the radial derivative and $V_{,\phi}$ is the $\phi$-derivative of the potential. Near a horizon we can expand 
\begin{align} 
f(r) &= f_1 (r-r_h) + \mathcal{O}(r-r_h)^2 \, , \\ 
N(r) &= N_0 + \mathcal{O}(r-r_h) \, .
\end{align}
Then, sufficiently close to any such horizon we can approximate the scalar equation by
\be 
\left(f_1 (r-r_h)\phi' \right)' = V_{,\phi}(\phi) \, .
\ee
Since $V_{,\phi}$ depends on $\phi$, assessing the behaviour of $\phi(r)$ near the horizon requires knowledge the potential. Here, for simplicity, we shall assume a massive scalar field with $V(\phi) = (\mu/2) \phi^2$. In that case, we can solve analytically for the behaviour of the scalar near the horizon:\footnote{We allow for the possibility that $\mu < 0$. This is permissible in anti de Sitter spacetimes, and since our analysis is local, it is ultimately insensitive to the asymptotic structure.}
\be 
\phi(r) = \phi_h + b_h \log \frac{|\mu| (r - r_h)}{|f_1|}   + \cdots \, , 
\ee
where $\phi_h$ and $b_h$ are constants of integration. We attach the subscript \(h\) to these quantities to emphasize that the analysis is local to a given horizon. If the spacetime contains multiple horizons, the same local form applies near each one, but with horizon-specific values of \(\phi_h\) and \(b_h\).

It is clear that when $b_h = 0$ we will have a (locally) regular solution. For this case it is easy to verify that the energy density and pressure are finite and the horizon regularity condition $p_r(r_h) = -\rho(r_h)$ is satisfied with
\be 
\rho(r_h) = V(\phi_h) = \frac{\mu \phi_h^2}{2} \, .
\ee
On the other hand, when $b_h \neq 0$ the scalar will be singular at the horizon. The energy density and pressure will diverge marginally,
\be\label{scalar_rho} 
\rho \sim p_r \sim \frac{f_1 b_h^2}{2 (r-r_h)}  \, .
\ee

To contextualize these findings, it is helpful to recall how one typically obtains black hole solutions sourced by a real scalar field $\phi(r)$. Such solutions can rarely be obtained analytically, and numerical methods must be used. The approach employed is to assume the scalar is regular at the event horizon, and obtain a power series solution for the scalar and metric functions there. The series expansion (for given values of the free parameters) is then used to prescribe initial data to a numerical integration scheme to integrate the solution either toward the asymptotic region or into the black hole interior. 

What this means for our local solution above is that at the event horizon one would set $b_h = 0$ to obtain a regular event horizon. However, it will generically not be possible to control the value of $b_h$ at more than one horizon in the spacetime. Hence, for the case of a real massive scalar, solutions with a regular event horizon will generally have $b_h \neq 0$ at any inner horizons. At such inner horizons, the scalar and its stress tensor components will diverge in the marginal manner studied above. 

Applying the results of our analysis we conclude that when considering quasi-topological gravity coupled to a real massive scalar, if the inverse function $H(x)$ decays polynomially, then the corresponding inner horizons will always be singular. However, if the inverse function decays \textit{super-polynomially} then our results indicate the inner horizon can remain regular. For example, for an inverse function that decays exponentially with $H'(S) \sim \exp(-a |S|)$, then the geometry will remain regular provided
\be 
\frac{4 \pi G_{\rm N}}{(D-2)} \frac{a |f_1| b_h^2}{r_h} > 1 \, .
\ee
Alternatively, if $H'(S) \sim \exp(-a |S|^\sigma)$ with $\sigma > 1$, then the geometry will generically be regular, requiring no further constraints on parameters. 

These conclusions are of course specific to the case of a real massive scalar. For different choices of potential, the degree of divergence will differ and may therefore lead to different conclusions for which types of resummation will yield regular or singular geometries. It would be an interesting task for future work to construct such solutions numerically ({\it e.g.}~with AdS asymptotics) and to classify the behaviour of more general scalar potentials.

\section{Electromagnetic quasi-topological gravities}\label{emqt}

We have so far focused on minimally coupled matter in QT gravities. However, as the gravitational sector contains an infinite tower of higher-curvature terms, the minimal coupling prescription is at best unnatural. In a more physically plausible setup, one would consider simultaneously higher-order corrections in the matter sector along with non-minimal couplings between curvature and matter. We now turn our attention to this problem. 

As an incipient endeavour in this direction, we will devote this section to the study of regular (both in matter and geometry) configurations in Electromagnetic quasi-topological (EMQT) gravities in $D \geq 4$~\cite{Cano:2020ezi,Cano:2020qhy,Cano:2022ord,Bueno:2022jbl}. These theories generalize purely gravitational QTs  with the inclusion of a non-minimally coupled $(D-3)$-form $\mathcal{A}$ which appears in the action via its field strength $\mathcal{F}= \mathrm{d}\mathcal{A}$, thus ensuring gauge invariance. More precisely, EMQTs are characterized by admitting static and spherically  symmetric black hole solutions~\eqref{Nf} with magnetic charge such that $N(r)=1$ and $f(r)$ is determined by a first-order differential equation which may be directly integrated into an algebraic equation, just like the usual QTs. There is an important difference, though: despite the fact that it is possible to define EMQTs to arbitrary order in the curvature and the gauge field strength, it is not clear whether these theories provide a basis for effective theories of gravity and a gauge $(D-3)$-form. Hence we should interpret the results in this section as a first foray into the interplay of matter and geometry in general effective theories of gravity and matter.

With these provisos in mind, let us consider the following theories of gravity and a $(D-3)$-form comprised of purely gravitational higher-curvature terms, higher-order matter terms and non-minimally coupled terms:\footnote{Note that here we are allowing for the possibility of finite or infinite towers of corrections in each sector, depending on whether one takes the terms $m_{\rm \max}$, $n_{\rm max}$, $m'_{\rm \max}$, $n'_{\rm max}$ to infinity or keeps some set of them finite.} 
\begin{widetext}
    \begin{align}
   \hspace{-0.2cm} \mathcal{L}&=R-\frac{\mathcal{F}^2}{(D-4)!}+ \sum_{m=2}^{m_{\rm max}} \frac{\widetilde{\lambda}_m}{((D-2)!)^m} \left ( \mathcal{F}^2 \right)^m +\sum_{n=2}^{n_{\rm max}} \widetilde{\alpha}_n \mathcal{Z}_n+\sum_{n=1}^{n'_{\rm max}} \sum_{m=1}^{m'_{\rm max}} \frac{\widetilde{\kappa}_{n,m}}{((D-2)!)^m} \mathcal{L}_{\rm EMQT}^{(n,m)}\,,\\ 
    \hspace{-0.2cm}  \mathcal{L}_{\rm EMQT}^{(n,m)}&=n  \left ( \mathcal{F}^2 \right )^{m-1} \left ( \mathcal{F}^2 \right )_{ab}^{cd} \left[ g_{n,m} \left (R^n \right)_{cd}^{ab}+  \left (R^{n-1} \right)_{cd}^{eb} \left ( R \, \delta_e^a+ 2 \, s_{n,m} R_e^a \right) +2(n-1) \left (R^{n-2} \right)_{ed}^{fb} R_c^e R_f^a \right]\,,
    \label{EMQTgs}
\end{align}
where we have made the following definitions:
\begin{align}
 \left (R^n \right)_{ab}^{cd}= &R_{ab}^{e_1 f_1} R_{e_1 f_1}^{e_2 f_2} \dots R_{e_{n-1} f_{n-1}}^{cd}\,,  \quad \left ( \mathcal{F}^2 \right )_{ab}^{cd}=\mathcal{F}_{ab e_1 \dots e_{D-4}} \mathcal{F}^{cd e_1 \dots e_{D-4}}\,, \quad \mathcal{F}^2=\left ( \mathcal{F}^2 \right )_{ab}^{ab}\,, \\  s_{n,m}&=4m+n-1-(2m+n-1)D\,, \quad g_{n,m}=\frac{(s_{n,m}+1)s_{n,m}}{2n}\,.
\end{align}
The couplings $\widetilde{\lambda}_m$ and $\widetilde{\kappa}_{n,m}$ are arbitrary, with dimensions of length$^{2(m-1)}$ and  length$^{2(m+n-1)}$, respectively. These theories comprise a special subclass of the two families of EMQTs identified in~\cite{Cano:2022ord,Bueno:2022jbl} which proves to be useful for our purposes. Throughout this section, we will consider the class of purely gravitational QTs to be contained in the EMQT family, arising in the limit of vanishing magnetic charge. 

\end{widetext}

\subsection{Limiting curvature hypothesis in electrovacuum}

Consider the equations of motion of the theory~\eqref{EMQTgs}, without further matter content (in \emph{electrovacuum}). We will be interested in static and spherically symmetric solutions~\eqref{Nf} endowed with a magnetic field strength:
\begin{equation}
    \mathcal{F}=P \, \mathrm{vol}_{S^{D-2}}\,,
\end{equation}
where $P$ is the magnetic charge and $\mathrm{vol}_{S^{D-2}}$ is the volume form of the $(D-2)$-dimensional round sphere. As it turns out, EMQTs of the form~\eqref{EMQTgs} admit spherically symmetric solutions determined by~\cite{Cano:2022ord,Bueno:2022jbl}
\begin{equation}
    N=1\,, \quad \frac{\mathrm{d}}{\mathrm{d}r} \left ( r^{D-1} h_y (\psi)\right)=0\,,
\end{equation}
where we have defined
\begin{align}
\hspace{-0.2cm}  h_y(\psi)&\equiv h_0(\psi)+h_y(0)+\sum_{n=1}^{n'_{\rm max}} \sum_{m=1}^{m'_{\rm max}} \kappa_{n,m} \psi^n y^m\,, \\
\hspace{-0.2cm}  h_0(\psi)& \equiv \psi+\sum_{n=2}^{n_{\rm max}} \alpha_n \psi^n\,, \\ 
\hspace{-0.2cm}   h_y(0)& \equiv y+\sum_{m=2}^{m_{\rm max}} \lambda_m y^m \,, \\ 
 \hspace{-0.2cm}    \kappa_{n,m}&\equiv\frac{(D-2n-2m(D-2))\, \tilde{\kappa}_{n,m}}{2^{-n+1}(D-2)}\,, \\
 \hspace{-0.2cm}    \lambda_{m}&\equiv\frac{ \tilde{\lambda}_{m}}{(D-2)(D-1-2m(D-2))}\,, \\
 \hspace{-0.2cm}    y&\equiv \frac{P^2}{r^{2(D-2)}}\,, \quad \psi \equiv \frac{1-f(r)}{r^2}\,.
\end{align}
Consequently, electrovacuum solutions are given by
\begin{equation}
    h_y(\psi)=S_0(r)\,, \quad S_0(r)=\frac{2\mathsf{M}}{r^{D-1}}\,,
    \label{eq:hEMQTg}
\end{equation}
where $\mathsf{M}$ is related to the mass $M$ of the solution as in~\eqref{eq:mass}. Restricting to positive mass $\mathsf{M}>0$ solutions, it is clear that $y=p^2 S_0^\sigma$ with $\sigma=2\dfrac{D-2}{D-1}$ and $(2\mathsf{M})^\sigma p^2=P^2$. Assuming $\partial_\psi h_y(\psi) \geq 0$---which is achieved, \emph{e.g.}, by imposing $\kappa_{n,m} \geq 0$---, it is possible to express $\psi$ in terms of $S_0$ in~\eqref{eq:hEMQTg} (at least implicitly), obtaining:
\begin{equation}
    \psi=H_p(S_0)\,,
\end{equation}
where we include the subscript $p$ to indicate that we are working with magnetically charged solutions. This expression is formally equivalent to the equation determining black hole solutions of purely gravitational QTs~\cite{Bueno:2024dgm}, except for the fact that $H_p(S_0)$ depends parametrically on the magnetic charge. As a result, expressions~\eqref{eq:R1f},~\eqref{eq:R2f} and~\eqref{eq:R3f} apply after exchanging $H(S)$ and derivatives thereof by $H_p(S_0)$ and its derivatives with respect to $S_0$ (denoted also with primes). Therefore, if 
\begin{equation}
    \left\lbrace H_p(S_0)\, , \,\,  S_0 H_p'(S_0)\, , \,\,  S_0^2 H_p''(S_0) \right\rbrace
    \label{eq:hipoEMQTg}
\end{equation}
remain bounded for $S_0 \in (0,+\infty)$ and $p^2 \in [0,+\infty)$, we conclude that all curvature invariants will be finite and (black hole) solutions will be regular.

\begin{figure}
    \centering
    \includegraphics[width=0.85\linewidth]{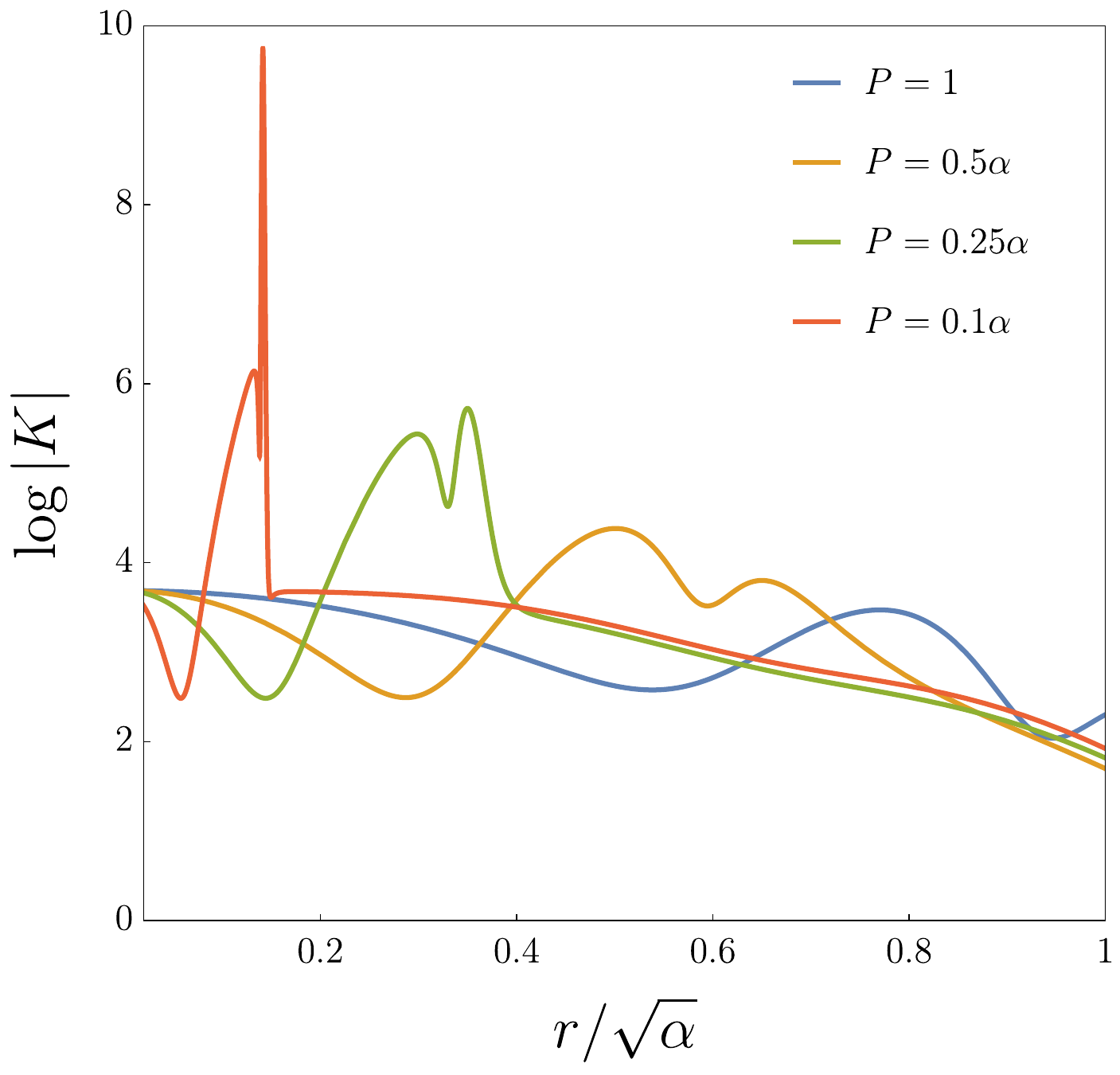}
    
    \caption{Logarithm of the Kretschmann scalar associated to various solutions of a five-dimensional EMQT with characteristic function $h_y^{\rm I}(\psi)$, as defined in~\eqref{eq:EMQTlch}. We explicitly observe that the Kretschmann scalar attains larger and larger values as $P \rightarrow 0$. Also, we see that the maximum appears for smaller values of $r$ as the charge diminishes. For the sake of concreteness, we have used $\kappa=\alpha=2\mathsf{M}$.}
    \label{fig:nlchEMQT}
\end{figure}

\begin{figure*}
    \centering
    \includegraphics[width=0.45\linewidth]{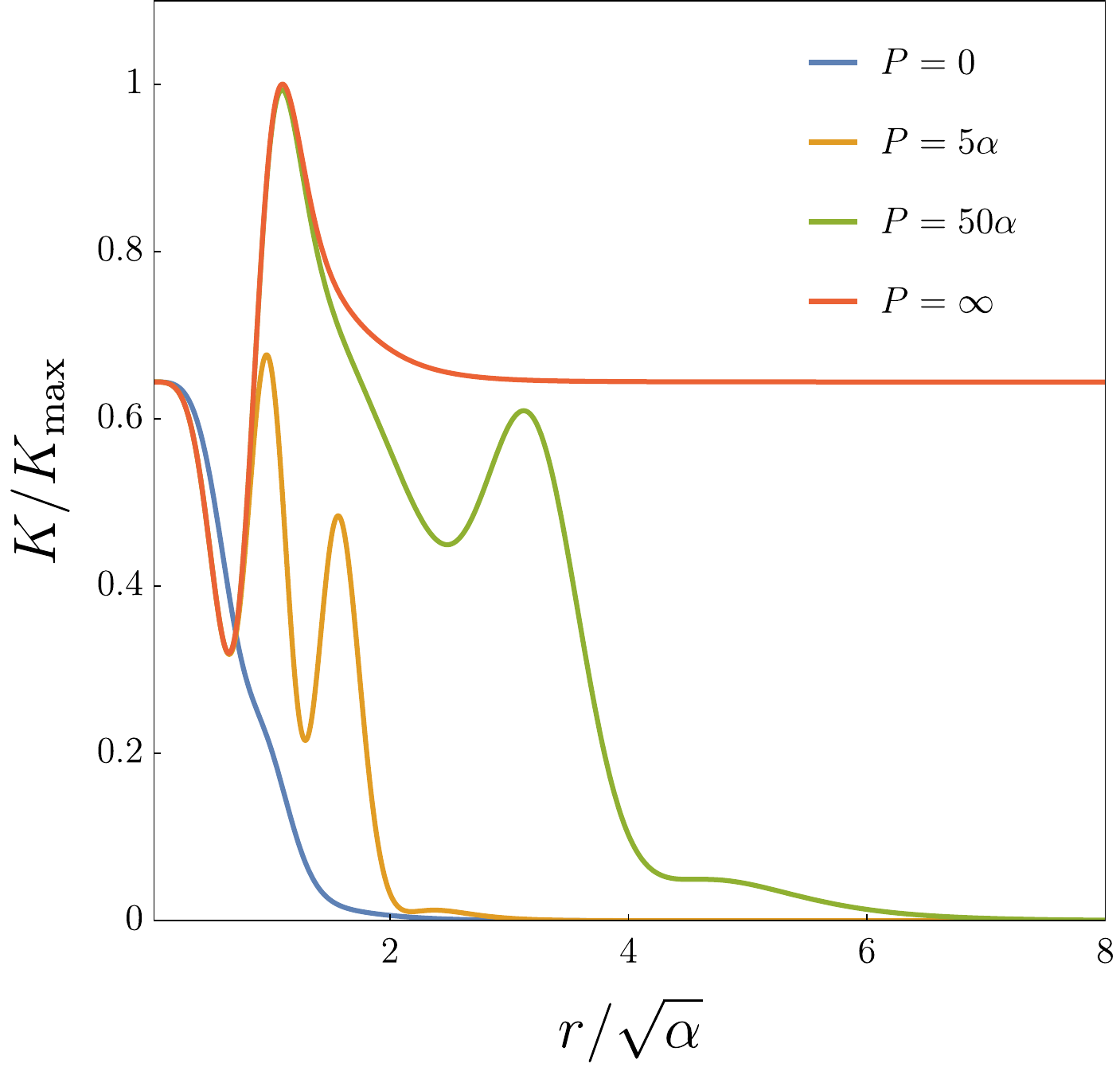}
    \includegraphics[width=0.45\linewidth]{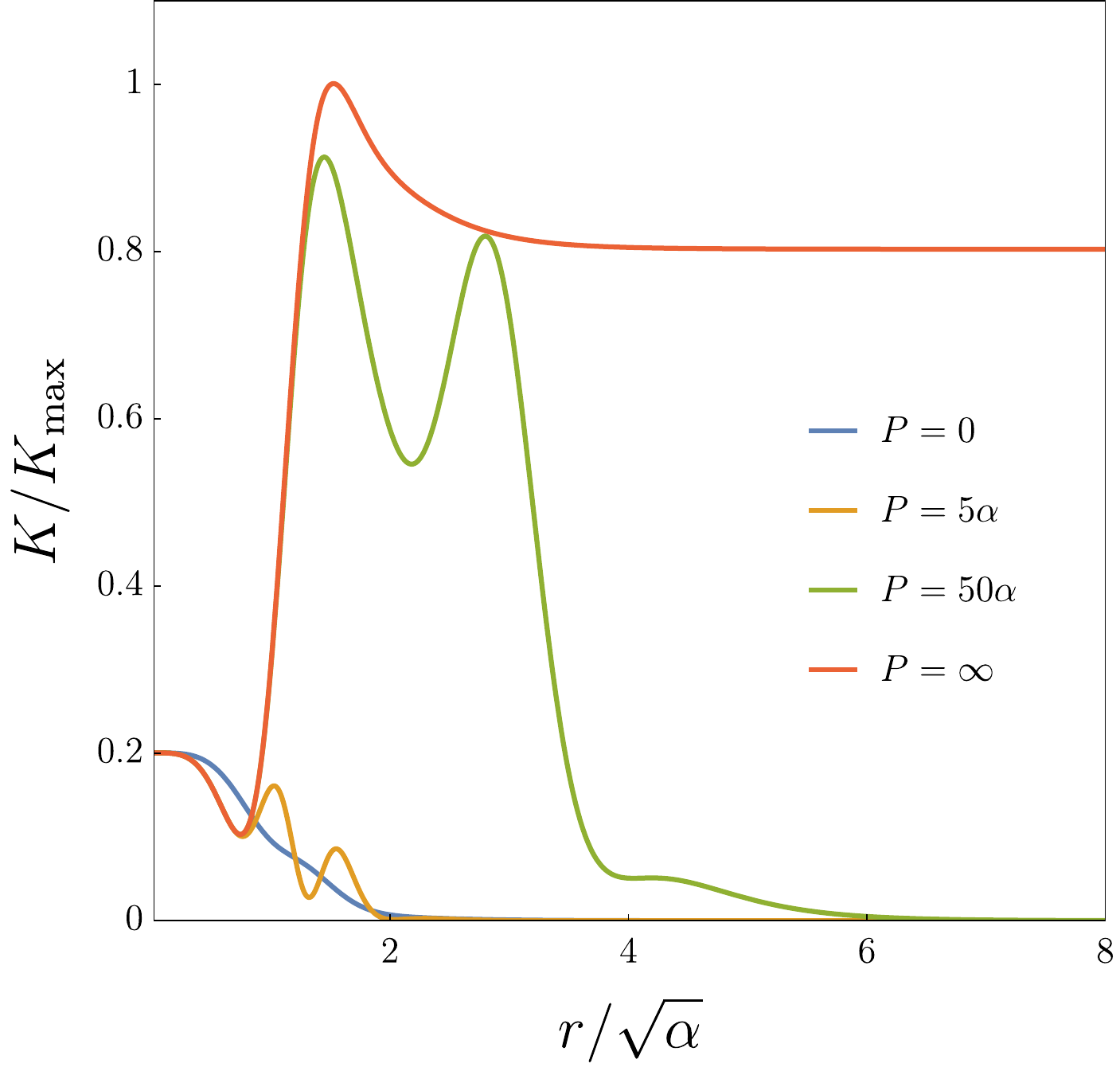}
    \caption{Kretschmann scalar associated to various solutions of a five-dimensional EMQT with characteristic function $h_y^{\rm II}(\psi)$, as defined in~\eqref{eq:ejemlcheq}. In the left plot, the maximum value of the Kretschmann scalar, $K_{\rm max}$, corresponds to $\sim 62/\alpha^2$, and we have used $\kappa=\mathsf{M}=\alpha$. In the right plot,  $K_{\rm max} \sim 200/\alpha^2$ and we have used $2\kappa=\mathsf{M}/3=\alpha$. We explicitly observe that the limiting curvature hypothesis is satisfied: curvature invariants are bounded by a universal quantity, independent of the mass and the charge. }
    \label{fig:lchEMQT}
\end{figure*}

A relevant question arises: will the limiting curvature hypothesis be satisfied by EMQTs with (magnetically charged) regular black hole solutions? By the assumptions~\eqref{eq:hipoEMQTg}, it turns out that curvature invariants will be bounded for every $p^2 \in [0,+\infty)$. However, this does not guarantee that the limiting curvature hypothesis will be satisfied, as one needs to show the existence of a universal bound (independent of the charge and mass) for \emph{any} curvature invariant. Specifically, the limiting curvature hypothesis will be satisfied if there exists $K \in \mathbb{R}$ independent of the mass and charge such that 
\begin{equation}
    K \geq \text{max}\left\lbrace H_p(S_0)\, , \,\,  S_0 H_p'(S_0)\, , \,\,  S_0^2 H_p''(S_0) \right\rbrace
    \label{eq:limhipoEMQTg}
\end{equation}
for every $ S_0 \in [0,\infty)$ and $p^2 \in [0,\infty)$. In the following, examples of theories with (magnetically charged) regular black holes that fulfill and do not fulfill the limiting curvature hypothesis are presented:
\begin{enumerate}
    \item {\bf RBHs without LCH}. Choose\footnote{In these examples, for the sake of simplicity, we will be considering theories whose uncharged solutions correspond to the $D$-dimensional Hayward black hole for odd $D \geq 5$~\cite{Bueno:2024dgm}. Nevertheless, other instances corresponding to different resummations of the purely gravitational sector may be found.} $\lambda_m=0$ and $\kappa_{n,m}=0$ if $m^2+n^2 >2$, $\kappa_{1,1}=\kappa > 0$ and $\alpha_n=\alpha^{n-1}$ for $\alpha>0$. In such a case, the characteristic function reads
    \begin{equation}
        h_y^{\rm I}(\psi)=\frac{\psi}{1-\alpha \psi}+y +\kappa y \, \psi\,.
        \label{eq:EMQTlch}
    \end{equation}
    Substituting in~\eqref{eq:hEMQTg}, solving for $\psi$ and selecting the solution smoothly connected to the Hayward black hole as $p^2 \rightarrow 0$, one finds
    \begin{equation}
        \psi=H_p^{\rm I}(S_0)=\frac{\tilde{S_0}-\sqrt{4p^2 S_0^{\sigma}(p^2 S_0^\sigma-S_0)\alpha \kappa+\tilde{S_0}^2}}{2p^2 \alpha \kappa S_0^{\sigma}}\,,
        \label{eq:1exEMQTg}
    \end{equation}
    where $\tilde{S_0}=1+\alpha S_0-p^2 S_0^\sigma(\alpha-\kappa)$. Observe that the radicand appearing in \eqref{eq:1exEMQTg} will always be nonnegative. For any finite $p$, it is easy to see that the terms~\eqref{eq:hipoEMQTg} for $H_p^{\rm I}$ will be bounded for every $S_0\in (0,+\infty)$ and $p^2 \in [0,+\infty)$. Hence the solutions always represent regular black holes. Nevertheless, this is not enough to guarantee that the limiting curvature hypothesis is satisfied. As a matter of fact, let us first focus on the limit
    of sufficiently large $p$, satisfying $p^2 S_0^\sigma \gg S_0$. In such a regime, \eqref{eq:1exEMQTg} simplifies to
    \begin{equation}
        H_p^{\rm I}= -\frac{1}{\kappa}+ \frac{1+S_0(\alpha+\kappa)}{\kappa(\alpha+\kappa) p^2 S_0^{\sigma}}+ \mathcal{O}(p^{-4})\,.
        \label{eq:H1ap1}
    \end{equation}
    On the other hand, let us now consider the vanishing charge limit. One directly finds
     \begin{equation}
       H_p^{\rm I} =  \frac{S_0}{1
    +\alpha S_0}-\frac{(1+(\alpha+\kappa) S_0) p^2 S_0^\sigma}{(1+\alpha S_0)^3}+\mathcal{O}(p^4)\,.
       \label{eq:H1ap2}
    \end{equation}
    Let us then examine the behaviour of $H_p^{\rm I}$ as $S_0 \rightarrow +\infty$ for sufficiently small magnetic charge. Indeed, if $S_0$ becomes large but $p^2 S_0^{\sigma}$ \emph{is comparatively small}---what indeed happens for small magnetic charge---the approximation~\eqref{eq:H1ap2} applies and will only transition into the regime governed by \eqref{eq:H1ap1} as $S_0$ is even larger. The crossover between these two regimes, taking place around the (nonzero) minimum value of the radicand in~\eqref{eq:1exEMQTg}, has drastic consequences on $S_0 \left ( H_p^{\rm I} \right)'$, which develops a steep jump near the transition as the magnetic charge goes to zero---we have  checked this numerically. As a result, $S_0^2 \left ( H_p^{\rm I} \right)''$ becomes arbitrarily large and the theory does not satisfy the limiting curvature hypothesis. This is illustrated in Figure~\ref{fig:nlchEMQT}.

   \item {\bf RBHs and LCH}. Take $\lambda_{2m}=0$, $\lambda_{2m+1}=(-1)^m \kappa^m$, $\alpha_n=\alpha^{n-1}$, $\kappa_{n,2m}=0$ and $\kappa_{n,2m+1}=\alpha^{n-1} (-1)^m \kappa^m$ for $\alpha, \kappa>0$. The characteristic function is
   \begin{equation}
       h_y^{\rm II}(\psi)=\frac{1}{1-\alpha \psi} \left ( \psi+ \frac{y}{1+\kappa y^2}\right)\,.
       \label{eq:ejemlcheq}
   \end{equation}
   We can invert analytically to obtain
   \begin{equation}
       \psi=H_p^{\rm II}(S_0)=\frac{S_0-p^2 S_0^\sigma(1-\kappa p^2 S_0^{1+\sigma})}{(1+\alpha S_0)(1+\kappa p^4 S_0^{2\sigma})}\,.
   \end{equation}
   It is straightforward to see that $H_p^{\rm II}(S_0)$ will give rise to finite curvature invariants for every $p\in [0,\infty)$. On the one hand, as the magnetic charge $p^2 \rightarrow +\infty$ one has
   \begin{align}
       H_p^{\rm II}(S_0)=\frac{S_0}{1+\alpha S_0}- \frac{1}{p^2 S_0^\sigma \kappa(1+\alpha S_0)}+ \mathcal{O}(p^{-6})\,,
       \label{eq:exIIdes1}
   \end{align}
   so the infinite-charge limit will give rise to bounded curvature invariants---amusingly, the solution tends to the \emph{neutral} black hole solution with the charge contributions effectively decoupling. Furthermore, the limit of vanishing magnetic charge is well defined:
   \begin{equation}
    H_p^{\rm II}(S_0)=\frac{S_0}{1+\alpha S_0}-\frac{p^2 S_0^\sigma}{1+\alpha S_0}+ \mathcal{O}(p^6)\,.
    \label{eq:exIIdes2}
   \end{equation}
   As the leading terms in~\eqref{eq:exIIdes1} and~\eqref{eq:exIIdes2} coincide, $H_p^{\rm II}$  does not feature the issues of the previous example as the magnetic charge goes to zero. As a result, there exists a universal bound for curvature invariants, independent of the charge and mass of the solution. This is explicitly shown in Figure~\ref{fig:lchEMQT}.
\end{enumerate}

We conclude this section with two remarks. First, we note that magnetically charged regular black holes can arise without the need for an infinite tower of higher-curvature terms, as long as the charge of the solution is \emph{nonzero}~\cite{Cano:2020ezi,Cano:2020qhy}. Indeed, take $\alpha_n=0$ for $n>2$, $\lambda_m=0$ for $m>2$ and $\kappa_{n,m}=0$ for every $n^2+m^2>2$, while $\kappa_{1,1}=1$. In this case, we trivially have that
\begin{equation}
    h_y^{\rm III}(\psi)=\psi+y+\psi y\,,
\end{equation}
while $\psi=H_p^{\rm III}(S_0)$ is obtained as
\begin{equation}
    H_p^{\rm III}(S_0)=\frac{S_0-p^2 S_0^\sigma}{1+p^2 S_0^\sigma}\,.
\end{equation}
As long as $p \neq 0$, the solution will have finite curvature invariants.

Secondly, one may ask about the case of Einstein gravity coupled to a non-linear theory of electrodynamics. In this latter case, the characteristic function adopts the form:
\begin{equation}
    h_y^{\rm IV}(\psi)=\psi+ Q(y)\,,
\end{equation}
where $Q(y)$ is a certain function of $y$. One has then
\begin{equation}
    \psi=H_p^{\rm IV}(S_0)=S_0-Q(p^2 S_0^\sigma)\,.
\end{equation}
Since $2\geq \sigma>1$ for every $D \geq 4$, no analytical choice for $Q(y)$ may render a bounded $H_p^{\rm IV}$. Furthermore, even if some non-analytical choice for $Q(y)$ is made so that a term proportional to $S_0$ in $Q(y)$ is generated, the cancellation of the divergence of $\psi$ as $S_0 \rightarrow \infty$ may only happen if the magnetic charge $P$ and the mass $\mathsf{M}$ are related. This is the prototypical case of  regular black hole models of GR coupled to a theory of non-linear electrodynamics~\cite{Ayon-Beato:1998hmi,Dymnikova:2004zc,Bronnikov:2000vy,Bronnikov:2022ofk}.

\section{Discussion}\label{disc}

In this paper we have studied how matter distorts the regularity of static and spherically symmetric spacetimes in QT gravities. The bulk of our results concerned minimally coupled matter, and a summary of our main findings can be found in Table~\ref{tab:matter-regularity}. We also performed a preliminary investigation of non-minimal coupling to matter. 

We established a largely model-independent understanding of when minimally coupled matter preserves or destroys the regularity of static, spherically symmetric solutions in QT gravities. The key object is
\begin{equation}
S(r)=\frac{2\mathsf{M}}{r^{D-1}}+\frac{2}{r^{D-1}}\int_\infty^r x^{D-2}\varrho(x)\,\mathrm{d}x\,,
\end{equation}
which sets the argument of the inverse characteristic function $H(S)$ through $\psi=H(S)$. Curvature singularities arise precisely when the matter sector diverges while $S(r)$ remains finite, so that $H'(S)$ and $H''(S)$ evaluate to ordinary finite numbers and cannot suppress the blow-up. Conversely, whenever a divergence in the matter sector forces $|S(r)|\to\infty$, the large-$|S|$ decay of $H'(S)$ and $H''(S)$ generically eliminates the would-be divergence in the curvature.

A central and somewhat counterintuitive consequence is that ``mild''  matter singularities (integrable finite-radius singularities in the energy density) are the most problematic cases insofar as regularity is concerned. By contrast, central $r = 0$ power-law divergences and non-integrable finite-radius divergences are generally compatible with bounded curvature, with the geometry remaining regular even when the stress tensor itself is singular (Table~\ref{tab:matter-regularity}). These conclusions require no equation of state and do not use matter field equations. They rely only on local asymptotic behaviour of $\rho,p_r,p_t$, and hence hold for general matter models.

In the presence of minimally coupled matter Markov's limiting curvature hypothesis no longer holds. Importantly, this failure is not synonymous with the appearance of a spacetime singularity. Rather, it means that the curvature scales of the geometry become solution-dependent ({\it e.g.} controlled by the mass, charge, or other parameters).

Our results also highlight the subtlety of coupling QT gravity to matter. Studying minimally coupled matter is convenient for pragmatic purposes: this setting provides a tractable arena in which to study deformations of the corresponding vacuum models while remaining largely agnostic about the particular details of the matter content. What is much less clear is how directly such results inform the fundamental interplay between matter and geometry in these theories. Recall that in five and higher dimensions the quasi-topological theories furnish a basis for the gravitational effective action in vacuum. From this perspective, incorporation of matter would generally proceed not by minimal coupling, but by including higher-derivative operators in the matter sector together with non-minimal couplings between matter and curvature. This is a substantially more difficult problem, and at present we do not know of any Einstein–matter effective field theories that can be mapped at all orders to a “quasi-topological-like” frame in which the equations of motion remain comparably accessible. 

Nevertheless, we performed a preliminary exploration of the effects of non-minimally coupled matter using EMQT gravities. These include non-minimal couplings to a $(D-3)$-form, may be defined at any order in the curvature and the gauge field strength, and generically allow for magnetically charged regular black hole solutions that are continuous deformations of the vacuum QT ones. We have presented explicit examples of  EMQTs in which the electrovacuum solutions are regular and all curvature invariants are bounded by mass- and charge-independent scales, {\it i.e.} Markov’s limiting curvature hypothesis holds. This underscores the limitation of conclusions based on minimally coupled matter: a singular minimal matter sector---or failure of Markov’s hypothesis there---need not reflect the true behaviour once the matter sector is resummed. Our result also shows that the limiting curvature hypothesis can serve as a criterion by which to select resummations of gravity plus matter actions.

Our results may have implications for the study of mass inflation in these models. Minimal matter couplings can be valuable as a proof of principle: the possibility of matter singularities yet regular geometry at Killing horizons suggests that QT gravity might evade mass-inflation–type instabilities, as recently argued in~\cite{Frolov:2026rcm}. At the same time, our broader conclusion—that coupling QT gravity to matter is subtle—tempers how definitive such analyses can be when they rely on minimal matter sectors. In this setting, a positive result (\emph{e.g.}, the absence of mass inflation in~\cite{Frolov:2026rcm}) is encouraging insofar as it demonstrates that avoiding the instability is possible \textit{in principle}. A negative result, however, is much less informative: as we have explicitly shown, higher-order matter corrections can play an essential role and could qualitatively change the conclusion. Overall, a conceptually complete treatment of mass inflation in QT gravity likely requires a framework that systematically incorporates higher-order matter terms.

There are a number of directions for future work. A natural extension would be to exploit the connection between QT gravities in spherical symmetry and two-dimensional Horndeski theories to consider the implications of (non)minimal matter for broader classes of theories. While QT gravities can be dimensionally reduced in spherical symmetry to a particular class of two-dimensional Horndeski theories~\cite{Bueno:2024zsx}, there is a much broader set of two-dimensional theories with similar properties~\cite{Carballo-Rubio:2025ntd,Borissova:2026krh}. It would be interesting to assess the generality of our results in that broader context, and moreover, to determine the necessary constraints (if any) for Markov's hypothesis to be realized in vacuum or otherwise in those models. In that setting, making contact with a broader class of theories than those defined by quasi-topological gravities requires considering terms non-polynomial in the curvature~\cite{Bueno:2025zaj,Borissova:2026wmn}. Such theories are on a more speculative footing than the polynomial quasi-topological actions, which can arise as perturbative corrections to general relativity within an effective field theory approach. Hence, limiting curvature may serve as a selection criterion to discriminate between different choices of non-polynomial gravitational actions, singling out representatives that may be better behaved.

\begin{table*}[p]
\begin{adjustwidth}{-0.05\textwidth}{-0.05\textwidth}
    \hspace{-0.02\textwidth}
    \renewcommand{\arraystretch}{1.15}
    \begin{tabular}{|p{\Sone}|p{\Stwo}|p{\Sthree}|p{\Sfour}|}
    \multicolumn{4}{c}{{\it \large Polynomial decay of the inverse function: $H'(S)\sim|S|^{-\beta}$, $H''(S)\sim|S|^{-(\beta+1)}$, with $\beta>1$}}
    \\ 
    \hline\hline
    \multirow[c]{1}{*}{\parbox{\Sone}{
    \textbf{Where}
    }}
    &
    \multirow[c]{1}{*}{\parbox{\Stwo}{
    \textbf{Local matter behaviour}
    }}
    &
    \multirow[c]{1}{*}{\parbox{\Sthree}{
    \textbf{Curvature outcome}
    }}
    &
    \multirow[c]{1}{*}{\parbox{\Sfour}{
    \textbf{Comments}
    }}
    \\
    \hline\hline
    \multirow[c]{1}{*}{\parbox{\Sone}{\vspace{0.50cm}
    Anywhere
    }}
    &
    Regular matter: $\vrho,\,\p_r,\,\p_t$ bounded (and derivatives entering $\mathcal{R}^{(i)}$ not singular)
    &
    \multirow[c]{1}{*}{\parbox{\Sthree}{\vspace{0.50cm}
    Regular geometry
    }}
    &
    Assuming the underlying vacuum theory already satisfies the bounded-$H$ conditions (so vacuum polynomial invariants are bounded)
    \\
    \hline
    \multirow[c]{3}{*}{\parbox{\Sone}{\vspace{1.5cm}
    $r\to 0^+$ \\ $f(0) = 1$
    }}
    &
    Central power-law density blow-up \newline 
    with $T_t^t = T_r^r \ \Longleftrightarrow \ \varrho=-\p_r$: \newline
    $\varrho \sim r^{-\gamma_1} \,, \quad \gamma_1>0$
    &
    \multirow[c]{1}{*}{\parbox{\Sthree}{\vspace{0.5cm}
    Regular geometry
    }}
    &
    \multirow[c]{1}{*}{\parbox{\Sfour}{\vspace{0.24cm}\raggedright
    $|S|\to\infty$ and the resummation suppresses the blow-up
    }}
    \\
    \cline{2-4}
    & 
    Central power-law density blow-up: \newline 
    $\varrho \sim r^{-\gamma_1} \,, \quad \gamma_1>0$ \newline
    $\p_r \sim r^{-\gamma_2}\,, \quad \gamma_2\in\mathbb{R}$
    &
    \multirow[c]{1}{*}{\parbox{\Sthree}{\vspace{0.24cm}
    Regular geometry \\
    iff pressure not too singular
    }}
    &
    Boundedness requires: \newline
    $\gamma_2\le \beta(D-1) \phantom{\,\gamma_1} \quad \text{if}\quad \gamma_1 \le D-1$ \newline 
    $\gamma_2\le \beta\,\gamma_1 \phantom{(D-1)} \quad \text{if}\quad \gamma_1 > D-1$ \\
    \cline{2-4} 
    & 
    Central pressure blow-up: \newline
    $\p_r \sim r^{-\gamma_2} \,, \quad \gamma_2>0$ \newline 
    with $\varrho(0^+)$ finite, {\it e.g.,} \newline
    $\varrho \sim r^{-\gamma_1} \,, \quad \gamma_1\leq0$
    &
    \multirow[c]{1}{*}{\parbox{\Sthree}{\vspace{0.65cm}
    Curvature singularity
    }}
    &
    \multirow[c]{1}{*}{\parbox{\Sfour}{\vspace{0.65cm}\raggedright
    $\p_rH'(S)$ diverges while $H'(S)$ is finite
    }}
    \\
    \hline
    \multirow{5}{*}{\parbox{\Sone}{\vspace{0.5cm}
    $r\to r_0$ \\ \vspace{0.3cm} s.t. either: \\ \vspace{0.3cm} $r_0=r_p^+>0$ \\ \vspace{0.3cm} or \\ \vspace{0.3cm}  $r_0=r_h^+>0$ \\ $f(r_h^+)=0^-$ \\ \vspace{0.1cm}(Inner or \\ cosmological \\ horizon)
    }}
    & 
    Non-integrable density blow-up \newline
    with $T_t^t = T_r^r \ \Longleftrightarrow \ \varrho=-\p_r$: \newline
    $\vrho\sim (r-r_0)^{-\gamma_1} \,, \quad \gamma_1>1$
    &
    \multirow[c]{1}{*}{\parbox{\Sthree}{\vspace{0.1cm}
    Regular geometry \\
    iff density and pressure \\ are  singular enough
    }}
    &
    \multirow[c]{1}{*}{\parbox{\Sfour}{\vspace{0.28cm}\raggedright
    Boundedness requires: \newline $\gamma_1\ge \frac{\beta+1}{\beta-1}$
    }}
    \\
    \cline{2-4}
    & 
    \vspace{0cm}\vspace{-0.15cm}
    Non-integrable density blow-up: \newline
    $\vrho\sim (r-r_0)^{-\gamma_1} \,, \quad \gamma_1>1$ \newline
    $\p_r\sim (r-r_0)^{-\gamma_2} \,, \quad \gamma_2\in\mathbb{R}$ \vspace{0.05cm}
    &
    \multirow[c]{1}{*}{\parbox{\Sthree}{\vspace{0.15cm}
    Regular geometry \\
    iff density is  singular enough \\
    and pressure not too singular
    }}
    &
    \multirow[c]{1}{*}{\parbox{\Sfour}{\raggedright\vspace{0.05cm}
    Boundedness requires: \\
    $\gamma_1\ge \frac{\beta+1}{\beta-1} \phantom{\text{ and }\gamma_2\leq\beta(\gamma_1-1)-1 \quad} \text{if} \quad  \gamma_2\le \gamma_1$ \\
    $\gamma_1\ge \frac{\beta+1}{\beta-1} \text{ and }\gamma_2\leq\beta(\gamma_1-1)-1 \quad \text{if} \quad  \gamma_2 > \gamma_1$
    }}
    \\
    \cline{2-4}
    &
    Marginal density blow-up: \newline
    $\vrho\sim (r-r_0)^{-\gamma_1} \,, \quad \gamma_1=1$
    &
    \multirow{3}{*}{\parbox{\Sthree}{\vspace{1.23cm}
    Curvature singularity
    }}
    &
    \parbox{\Sfour}{\vspace{0.24cm}\raggedright
    Matter fields grow exponentially in $|S|$
    }
    \\
    \cline{2-2} \cline{4-4}
    & 
    Integrable density blow-up, {\it e.g.,} \newline $\vrho\sim (r-r_0)^{-\gamma_1} \,, \quad 0<\gamma_1<1$
    &
    &
    Terms such as $\vrho\,H'(S)$, $r\vrho' H'(S)$ diverge while $H'(S)$ is finite
    \\
    \cline{2-2} \cline{4-4}
    &
    Pressure blow-up: \newline
    $\p_r \sim (r-r_0)^{-\gamma_2} \,, \quad \gamma_2>0$ \newline 
    with $\varrho(r_0)$ finite or integrable
    &
    &
    \multirow[c]{1}{*}{\parbox{\Sfour}{\vspace{0.4cm}\raggedright
    $\p_rH'(S)$ diverges while $H'(S)$ is finite
    }}
    \\
    \hline
    \parbox{\Sone}{\vspace{0.1cm}
    $r\to r_h^+$ \\ 
    $f(r_h^+)=0^\pm$ 
    \vspace{0.1cm}
    }
    &
    \parbox{\Stwo}{\vspace{0.1cm}\raggedright
    Static matter at a Killing horizon \\
    with finite $\varrho$ and $\p_r$
    }
    &
    \parbox{\Sthree}{\vspace{0.1cm}
    Curvature singularity \\ 
    due to ``blueshift'' divergence
    }
    &
    \parbox{\Sfour}{\vspace{0.1cm}\raggedright
    Avoided if the horizon regularity condition holds: $T_t^t(r_h)=T_r^r(r_h) \ \Longleftrightarrow \ \p_r(r_h)=-\vrho(r_h)$
    }
    \\
    \hline\hline
    \multicolumn{4}{c}{\vspace{0.0cm}}
    \\ 
    \multicolumn{4}{c}{{\it \large Faster than polynomial decay of the inverse function: $H'(S)\sim e^{-a|S|}$, with $a>0$}}
    \\ 
    \hline
    \multicolumn{4}{l}{ Main differences between polynomial (LHS of the arrows) and faster than polynomial decays (RHS of the arrows):}
    \\ 
    \multicolumn{4}{c}{\vspace{0.25cm}
    \begin{tabular}{c c c}
        Regular geometry 
        & $\Longrightarrow$ 
        & Regular geometry \\
        
        Regular geometry under some conditions 
        & $\Longrightarrow$ 
        & Regular geometry \\
        
        Curvature singularity due to exponential growth of matter fields 
        & $\Longrightarrow$ 
        & Regular geometry under some conditions (see below) \\
        
        Other types of curvature singularities
        & $\Longrightarrow$ 
        & Curvature singularities
    \end{tabular}
    }
    \\ 
    \hline\hline 
    \parbox{\Sone}{
    \textbf{Where}
    }
    &
    \parbox{\Stwo}{
    \textbf{Local matter behaviour}
    }
    &
    \parbox{\Sthree}{
    \textbf{Curvature outcome}
    }
    &
    \parbox{\Sfour}{
    \textbf{Comments}
    }
    \\
    \hline\hline
    \parbox{\Sone}{\vspace{0.1cm}
    $r\to r_0$ \\ \vspace{0.1cm} s.t. either \\ \vspace{0.1cm} $r_0=r_p^+>0$ \\ \vspace{0.1cm} or \\ \vspace{0.1cm}  $r_0=r_{h}^+$ \\ $f(r_h^+)=0^-$ \vspace{0.1cm}
    }
    &
    \parbox{\Stwo}{\vspace{0.15cm}\raggedright
    Marginal density blow-up: \newline
    $\vrho\sim (r-r_0)^{-\gamma_1} \,, \quad \gamma_1=1$ \newline
    $\p_r\sim (r-r_0)^{-\gamma_2} \,, \quad \gamma_2\in\mathbb{R}$ \vspace{0.1cm}\newline
    $S(r)\sim\lambda_{\rm log} \log (r-r_0)$ where \\ 
    $\lambda_{\rm log}=\frac{2}{r_0}\lim_{r\to r_0}(r-r_0)\varrho(r)$
    }
    &
    \parbox{\Sthree}{\vspace{0.15cm}
    Regular geometry \\ 
    iff the geometry decays fast enough
    }
    &
    \parbox{\Sfour}{\vspace{0.15cm}\raggedright
    The exponential behaviour of the matter fields is suppressed by the geometry invariants if \newline
    $a|\lambda_{\rm log}| \geq \max\{2,1+\gamma_2\}$
    }
    \\ 
    \hline  \hline   
    \end{tabular}
    \end{adjustwidth}
    \caption{Summary of results for minimally coupled matter. We describe the different scenarios studied in this work, expressing for each matter configuration the resulting curvature behaviour along with sufficient conditions to ensure that outcome. The theories are defined by an invertible characteristic function $h(\psi)$ with $\psi=H(S)=h^{-1}(S)$. The parameter $\beta>1$ is defined by the large-$|S|$ decay $H'(S)\sim |S|^{-\beta}$ and $H''(S)\sim |S|^{-(\beta+1)}$ that is required to ensure boundedness of vacuum curvature invariants. For faster-than-polynomial decays of $H'(S)$, the geometry remains regular for a broader class of singular matter distributions than in the polynomial decay case.}
    \label{tab:matter-regularity}
\end{table*}

Concerning non-minimal matter models, it would be particularly interesting to determine whether EMQTs are sufficiently general to capture electrovacuum effective field theory. While this is at present unclear,  these models could serve, at the very least, as useful proxies towards the systematic study of the interplay between matter and geometry in general effective theories. Interestingly enough, for the class of EMQTs we examined, we observed that these lead to (magnetically charged) regular black hole solutions under analogous conditions to those of purely gravitational QTs. However, differently from the latter, ensuring the existence of regular black holes does not guarantee the fulfillment 
 of Markov's limiting curvature hypothesis in electrovacuum. Rather than a drawback, this provides us with a powerful physical criterion for further refining the class of EMQTs.

Another important question concerns the role of curvature invariants involving covariant derivatives of curvature---see \cite{Borissova:2026krh}. Even in vacuum these will only be bounded in the first place provided that the metric function under consideration is $C^\infty$ near $r = 0$. More concretely, this translates to the requirement that, should the metric function admit a series expansion in the vicinity of $r = 0$, then that series expansion should have no odd powers of $r$~\cite{Giacchini:2021pmr}. For example, this holds for the Hayward black holes provided that $D$ is odd. Here we note that curvature invariants featuring covariant derivatives are still bounded for $C^\infty$ regular black holes, but the bound is not independent of the mass. This is because the covariant derivative introduces new objects which cannot be written in terms of just $S$ and (derivatives of) $H(S)$, and hence the scaling property of the polynomial invariants is lost. In the case of the quasi-topological theories we have studied here the issue is not so important as the actions of the theories we consider do not feature covariant derivatives, and hence both the solutions and the actions are bounded in a mass-independent manner. Nonetheless, it would be interesting---though more involved---to study the boundedness of invariants containing covariant derivatives of curvature. It would be especially interesting to identify a regular black hole metric for which \textit{all} curvature invariants have a mass independent bound---or to prove that such a construction is impossible.

\section*{Acknowledgements}
We thank  Pablo A. Cano and Pedro G.~S.~Fernandes for useful discussions. PB was supported by a Proyecto de Consolidación Investigadora (CNS 2023-143822) from Spain’s Ministry of Science, Innovation and Universities, and by the grant PID2022-136224NB-C22, funded by MCIN/AEI/ 10.13039/501100011033/FEDER, UE.
\'AJM  was supported by a Juan de la Cierva contract (JDC2023-050770-I) from Spain’s Ministry of Science, Innovation and Universities. 
AVC was supported by a Spanish Government Fellowship FPU24/00899. 
The authors also acknowledge financial support from the grant CEX2024-001451-M funded by MICIU/AEI/10.13039/501100011033. 

\appendix 

\section{Regularity for power-law matter blow-ups}
\label{app:blowups_unified}

This appendix collects the regularity estimates for stress tensors that develop
power-law divergences either at the center \(r=0\) or at a finite radius \(r=r_p\).
The two cases discussed in the main text are recovered as special cases:
\begin{itemize}
\item \(T^t_t=T^r_r\): the anisotropy function \(W(r)\) vanishes and the
analysis reduces to bounds involving only \(\vrho\).
\item \(T^t{}_t\neq T^r{}_r\): \(W(r)\) contributes additional terms, leading to extra
inequalities involving the pressure falloff exponent.
\end{itemize}

Throughout we assume \(D\ge 4\) and use the same definitions as in the main text. We assume that $f(r)$ is nonzero at the location of any matter blow up---the case where the matter blows up at a Killing horizon was treated in the main text. The potentially dangerous matter-sourced contributions to curvature invariants arise as combinations of the terms 
\begin{align}
&\rho\,H'(S),\quad r\rho'\,H'(S),\quad \rho^2\,H''(S),\quad \rho\,S\,H''(S),
\nonumber
\\
&p_r\,H'(S), \quad 
rW(r)\,H(S), \quad rW(r)\,\rho H'(S)\, ,
\nonumber
\\
&rW(r)\,S H'(S) \, , \quad W'(r) \, , \quad W^2(r)\,.
\label{eq:dangerous_terms}
\end{align}

We assume the same large-\(|S|\) decay that is sufficient to bound the \emph{vacuum}
curvature invariants:
\begin{equation}
H'(S)\sim \frac{1}{|S|^{\beta}}\,,\quad
H''(S)\sim \frac{1}{|S|^{\beta+1}}\,,
\quad |S|\to\infty\,,
\label{eq:H_decay_beta}
\end{equation}
with \(\beta>1\). Of course, some resummations have $H(S)$ decay faster than any polynomial. In these cases, sometimes less restrictive conditions than what we will find below will hold. However, it is more difficult to assess all such possibilities in a coherent manner. We therefore stress that for resummations with faster than polynomial decays, the following results will correspond to sufficient, but not necessary, conditions.

\subsection*{1. Central power-law blow-up at \texorpdfstring{\(r=0\)}{r=0}}

Assume that near the origin the energy density diverges as
\begin{equation}
\vrho(r)\sim \frac{1}{r^{\gamma_1}},\qquad r\to 0^+\, ,
\label{eq:rho_center}
\end{equation}
 with $\gamma_1 >0$ and  \(\vrho\) otherwise smooth and such that the integral defining \(S(r)\) is
well-behaved away from \(r=0\). In the anisotropic case we also allow
\begin{equation}
\p_r(r)\sim \frac{1}{r^{\gamma_2}} \, ,\qquad r\to 0^+ \, ,
\label{eq:pr_center}
\end{equation}
for some exponent $\gamma_2$, while \(W(r)\) is the same
anisotropy quantity used in the main text (so that \(W= 0\) when \(T^t_t=T^r_r\)).

We can then isolate the divergent part of \(S(r)\) at small \(r\):
\begin{equation}
S(r)\sim \frac{2\mathsf{M}}{r^{D-1}}-\frac{\lambda}{r^{\gamma_1}}+\text{finite} \, ,
\qquad r\to 0^+ \, ,
\label{eq:S_center_asymp}
\end{equation}
for some constant \(\lambda\). We then consider two possibilities in turn.

\subsubsection*{1.A Matter-dominated center: \texorpdfstring{\(\gamma_1>D-1\)}{gamma\_1>D-1}}

When $\gamma_1 > D-1$, the matter term dominates \(S\) at small \(r\), and one may invert
\begin{equation}
r\sim \Big(\frac{|\lambda|}{|S|}\Big)^{1/\gamma_1}\,.
\label{eq:r_of_S_matterdom}
\end{equation}
Then
\begin{equation}
\vrho \sim |S| \, ,
\qquad r\vrho'\sim |S| \, ,
\qquad
\p_r\sim |S|^{\gamma_2/\gamma_1} \, .
\label{eq:scalings_matterdom}
\end{equation}
Using~\eqref{eq:H_decay_beta}, the density-driven terms in~\eqref{eq:dangerous_terms}
scale as
\begin{align}
\vrho H'(S)&\sim |S|^{1-\beta} \, ,\quad
\\
r\vrho' H'(S)&\sim |S|^{1-\beta} \, ,\quad
\\
\vrho^2H''(S)&\sim |S|^{1-\beta} \, ,\quad
\\
\vrho S H''(S)&\sim |S|^{1-\beta} \, ,
\label{eq:basic_terms_matterdom}
\end{align}
which are bounded for \(\beta>1\). If $T_t^t = T_r^r$, then the argument concludes at this stage. 

If $T_t^t \neq T_r^r$, a potentially new requirement comes from the pressure sector:
\begin{equation}
\p_rH'(S)\sim |S|^{\gamma_2/\gamma_1-\beta} \, ,
\label{eq:pr_term_matterdom}
\end{equation}
so \(\p_rH'(S)\) is bounded provided
\begin{equation}
\gamma_2\le\beta\gamma_1 \, .
\label{eq:cond_gamma2_matterdom}
\end{equation}
If $T_t^t \neq T_r^r$, terms involving \(W(r)\) appear (cf.~\eqref{eq:dangerous_terms}). Given that the terms involving the energy density are bounded, several of the terms involving $W(r)$ simplify and 
we need to require only the boundedness of 
\be 
W(r) \quad \text{and} \quad W'(r) \, .
\ee
We can work out these terms to find that
\begin{align}
W(r) &\sim C_1|S|^{\frac{-1 + (1-\beta) \gamma_1}{\gamma_1}} + C_2 |S|^{ \frac{-1 + \gamma_2 - \beta\gamma_1}{\gamma_1}} \, ,
\\
W'(r) & \sim C_3 |S|^{1-\beta} + C_4 |S|^\frac{\gamma_2 - \beta \gamma_1}{\gamma_1} \, .
\end{align}
If Eq.~\eqref{eq:cond_gamma2_matterdom} holds, then the decay of these terms introduce no new requirements. 

\subsubsection*{1.B Vacuum-dominated center: \texorpdfstring{\(\gamma_1<D-1\)}{gamma\_1<D-1}}
Now the mass term dominates \(S\), giving
\begin{equation}
r\sim \Big(\frac{2\mathsf{M}}{S}\Big)^{1/(D-1)} \, .
\label{eq:r_of_S_vacdom}
\end{equation}
Hence
\begin{equation}
\vrho\sim S^\frac{\gamma_1}{(D-1)} \, ,
\!\quad r\vrho'\sim S^\frac{\gamma_1}{(D-1)}\, ,
\!\quad \p_r\sim S^\frac{\gamma_2}{(D-1)} \, .
\label{eq:scalings_vacdom}
\end{equation}
Then the density-driven terms behave as
\begin{align}
\vrho H'(S) &\sim S^{\frac{\gamma_1}{(D-1)}-\beta} \, , \\
r\vrho' H'(S) &\sim S^{\frac{\gamma_1}{(D-1)}-\beta} \, , \\
\vrho^2 H''(S) &\sim S^{\frac{2\gamma_1}{(D-1)}-\beta-1} \, ,\\
\vrho S H''(S) &\sim S^{\frac{\gamma_1}{(D-1)}-\beta} \, ,
\end{align}
which are bounded under the same vacuum requirement \(\beta>1\) because
\(\gamma_1<D-1\) implies \(\gamma_1/(D-1)<1\). If $T_t^t = T_r^r$, then the argument concludes here: the vacuum boundedness conditions ensure the boundedness in the presence of matter with a central power law singularity.

For the pressure term one has
\be
\p_rH'(S)\sim S^{\frac{\gamma_2}{(D-1)}-\beta} \, ,
\ee
so boundedness is ensured if
\begin{equation}
\gamma_2\le\beta(D-1) \, .
\label{eq:cond_gamma2_vacdom}
\end{equation}
As in Case A.1, many of the $W$-dependent terms simplify upon using the boundedness of the above density-driven terms. The conditions reduce to requiring boundedness of 
\be 
W(r) \quad \text{and} \quad W'(r) \, ,
\ee
which in this case work out to be
\begin{align}
    W(r) &\sim C_1 S^{\frac{\gamma_1 -1}{D-1} - \beta} + C_2 S^{\frac{\gamma_2-1}{D-1} - \beta} \, , 
    \\
    W'(r) &\sim 
     C_3 S^{\frac{\gamma_1}{D-1} - \beta} + C_4  S^{\frac{\gamma_2}{D-1}-\beta} \, .
\end{align}
There are no further conditions beyond Eq.~\eqref{eq:cond_gamma2_vacdom}.

\subsubsection*{1.C Marginal central blow-up: \texorpdfstring{$\gamma_1 = D-1$}{gamma\_1=D-1}}

In this section we treat the marginal case for a central divergence,
\(\gamma_1=D-1\), which is not covered by Cases 1.A--1.B.  Assume
\begin{equation}
\vrho(r)\sim \frac{A}{r^{D-1}} \, ,\qquad r\to 0^+ \, ,
\end{equation}
and in the anisotropic case allow for 
\be 
\p_r(r)\sim \frac{B}{r^{\gamma_2}}
\ee 
as in~\eqref{eq:pr_center}.
Then the matter integrand in \(S(r)\) behaves as \(x^{D-2}\vrho(x)\sim A/x\), so the
matter integral develops a logarithmic divergence. More precisely, for any fixed
\(r_0>0\) one may write
\begin{align}
&\int_{\infty}^{r}x^{D-2}\vrho(x)\,\dif x
=
\int_{\infty}^{r_0}x^{D-2}\vrho(x)\,\dif x
\nonumber \\
&+\int_{r_0}^{r}\left(x^{D-2}\vrho(x)-\frac{A}{x}\right)\,\dif x
+A\int_{r_0}^{r}\frac{\dif x}{x}
\nonumber\\
&\equiv C + A\log\!\left(\frac{r}{r_0}\right)\, ,
\qquad r\to 0^+ \, ,
\label{eq:I_center_marginal}
\end{align}
where \(C\) is finite by construction. Substituting into the definition of $S$ gives
\begin{align}
S(r)&\sim \frac{2\mathsf{M}+2C+2A\log\!\left(\frac{r}{r_0}\right)}{r^{D-1}}
\nonumber 
\\
&\equiv \frac{2\mathsf{M}_{\rm eff}+2A\log\!\left(\frac{r}{r_0}\right)}{r^{D-1}}\,,
\qquad r\to 0^+ \, ,
\label{eq:S_center_marginal}
\end{align}
so that \(|S|\to\infty\) and, for sufficiently small \(r\),
\begin{equation}
|S(r)|\sim \frac{2|A|\,|\log r|}{r^{D-1}} \,.
\label{eq:S_center_marginal_dom}
\end{equation}
The large-\(|S|\) decay~\eqref{eq:H_decay_beta} then implies the useful small-\(r\) estimates
\begin{equation}
H'(S)\sim \frac{r^{\beta(D-1)}}{|\log r|^{\beta}} \,,
\
H''(S)\sim \frac{r^{(\beta+1)(D-1)}}{|\log r|^{\beta+1}}\,,
\ r\to 0^+\,.
\label{eq:Hprime_center_marginal}
\end{equation}

Using \(\vrho\sim r^{-(D-1)}\) and \(r\vrho'\sim-(D-1)\vrho\), the density-driven terms in
\eqref{eq:dangerous_terms} scale as
\begin{align}
\vrho H'(S)&\sim \frac{r^{(\beta-1)(D-1)}}{|\log r|^{\beta}}\,,
\\
r\vrho' H'(S)&\sim \frac{r^{(\beta-1)(D-1)}}{|\log r|^{\beta}}\,,
\\
\vrho^2H''(S)&\sim \frac{r^{(\beta-1)(D-1)}}{|\log r|^{\beta+1}}\,,
\\
\vrho S H''(S)&\sim \frac{r^{(\beta-1)(D-1)}}{|\log r|^{\beta}}\,,
\label{eq:basic_terms_marginal}
\end{align}
which are bounded (in fact they vanish) as \(r\to 0^+\) provided \(\beta>1\). Hence, when
\(T^t_t=T^r_r\) (so \(W= 0\)), the marginal blow-up \(\gamma_1=D-1\) does not cause curvature singularities provided there are none already in the vacuum case. 

If \(T^t{}_t\neq T^r{}_r\), the pressure sector contributes
\begin{equation}
\p_r H'(S)\sim \frac{r^{\beta(D-1)-\gamma_2}}{|\log r|^{\beta}}\,,
\label{eq:pr_term_marginal}
\end{equation}
so \(\p_rH'(S)\) is bounded provided
\begin{equation}
\gamma_2\le \beta(D-1)\,.
\label{eq:cond_gamma2_marginal}
\end{equation}
Assuming the boundedness of the density-driven terms~\eqref{eq:basic_terms_marginal}, the remaining \(W\)-dependent contributions in~\eqref{eq:dangerous_terms} reduce, as in Cases
A.1--A.2, to requiring boundedness of \(W(r)\) and \(W'(r)\). Using the same scaling that led to
\eqref{eq:basic_terms_marginal}, their dominant contributions behave as
\begin{align}
W(r) &\sim
C_1\,\frac{r^{(\beta-1)(D-1)+1}}{|\log r|^{\beta}}
+ C_2\,\frac{r^{\beta(D-1)+1-\gamma_2}}{|\log r|^{\beta}} \,,
\label{eq:W_marginal}
\\
W'(r) &\sim
C_3\,\frac{r^{(\beta-1)(D-1)}}{|\log r|^{\beta}}
+ C_4\,\frac{r^{\beta(D-1)-\gamma_2}}{|\log r|^{\beta}}\,.
\label{eq:Wprime_marginal}
\end{align}
The first terms in~\eqref{eq:W_marginal}--\eqref{eq:Wprime_marginal} are bounded for \(\beta>1\).
The second term in \(W'(r)\) is bounded precisely under~\eqref{eq:cond_gamma2_marginal}, and
no further restrictions arise. In particular, the marginal case \(\gamma_1=D-1\) yields the same
pressure bound as the vacuum-dominated branch~\eqref{eq:cond_gamma2_vacdom}.

\subsubsection*{Summary of central power-law blow-ups}
For \(T^t{}_t= T^r{}_r\), a central power-law blow-up of the form \(\rho\sim r^{-\gamma_1}\) does not force curvature
divergences provided vacuum curvature is bounded (\(\beta>1\)). If \(T^t{}_t\neq T^r{}_r\),
one additionally requires the pressure exponent to satisfy
\be 
\begin{cases}
\gamma_2 \le \beta (D-1) \quad &\text{if} \quad \gamma_1 \le D-1  \, ,
\\
\gamma_2 \le \beta \gamma_1 \quad &\text{if} \quad \gamma_1 > D-1 \, .
\end{cases}
\ee

\subsection*{2 Finite-radius blow-up at \texorpdfstring{\(r=r_p\)}{r=r\_p}}

\subsubsection*{2.A Power law blow-up with \texorpdfstring{\(\gamma_1 > 1\)}{gamma\_1>1}}

Assume the energy density diverges at some finite radius \(r_p>0\) as
\begin{equation}
\vrho(r)\sim \frac{A}{(r-r_p)^{\gamma_1}},\qquad r\to r_p^+,
\label{eq:rho_rp}
\end{equation}
with \(\gamma_1>0\), while \(\rho\) is otherwise smooth for \(r>r_p\) and the integral defining
\(S(r)\) converges. In the anisotropic case we also allow
\begin{equation}
\p_r(r)\sim \frac{B}{(r-r_p)^{\gamma_2}},\qquad r\to r_p^+,
\label{eq:pr_rp}
\end{equation}
for some exponent \(\gamma_2\) and constant \(B\), while \(W(r)\) is the same
anisotropy quantity used in the main text (so that \(W= 0\) when \(T^t_t=T^r_r\)).

The most singular piece of the matter integral
\begin{equation}
I(r)=\int_{\infty}^{r}x^{D-2}\vrho(x)\,\dif x
\end{equation}
behaves, for \(\gamma_1\neq 1\), as
\begin{equation}
I(r)\sim -\frac{\,r_p^{D-2}A}{(\gamma_1-1)(r-r_p)^{\gamma_1-1}} \, ,
\quad r\to r_p^+,
\label{eq:I_rp_asymp}
\end{equation}
and therefore
\begin{equation}
S(r)\sim \frac{\lambda}{(r-r_p)^{\gamma_1-1}}
\label{eq:S_rp_asymp}
\end{equation}
for some constant \(\lambda\).

If \(0<\gamma_1<1\), the integral \(I(r)\) stays finite as \(r\to r_p\), so \(S(r)\) does not
run to \(\pm\infty\) and the resummation cannot suppress the matter blow-up.  The curvature invariants diverge in this case, as discussed in the main text. Henceforth let us assume $\gamma_1 > 1$. 

If $\gamma_1 > 1$, then \(|S|\to\infty\) as \(r\to r_p\), and one can trade \((r-r_p)\) for \(|S|\):
\begin{equation}
(r-r_p)\sim |S|^{-1/(\gamma_1-1)}.
\end{equation}
Then
\begin{equation}
\vrho\sim |S|^{\frac{\gamma_1}{(\gamma_1-1)}},
\quad
r\vrho' \sim |S|^{\frac{\gamma_1+1}{(\gamma_1-1)}},
\quad 
\p_r \sim |S|^{\frac{\gamma_2}{(\gamma_1-1)}} \, . 
\label{eq:scalings_rp}
\end{equation}
The potentially dangerous terms involving the density scale as
\begin{align}
\vrho H'(S) &\sim |S|^{\frac{\gamma_1}{(\gamma_1-1)}-\beta},\\
r\vrho' H'(S) &\sim |S|^{\frac{\gamma_1+1}{(\gamma_1-1)}-\beta},\\
\vrho^2 H''(S) &\sim |S|^{\frac{2\gamma_1}{(\gamma_1-1)}-(\beta+1)},\\
\vrho S H''(S) &\sim |S|^{\frac{\gamma_1}{(\gamma_1-1)}-\beta}.
\end{align}
All of these are bounded provided
\begin{equation}
\gamma_1 \ge \frac{\beta+1}{\beta-1}.
\label{eq:finite_r_condition}
\end{equation}

For the pressure term one has
\be 
\p_r H'(S) \sim |S|^{\frac{\gamma_2}{(\gamma_1-1)}-\beta}
\ee
which is bounded provided that $\gamma_2 \le \beta (\gamma_1 - 1)$. Assuming the boundedness of the terms involving the density, various terms involving $W(r)$ further simplify and the problem reduces to the boundedness of $W(r)$ and $W'(r)$. For the present case, the dominant contributions can be worked out to be
\begin{align}
W(r) &\sim C_1 |S|^{\frac{\gamma_1}{\gamma_1-1}-\beta} + C_2 |S|^{\frac{\gamma_2}{\gamma_1-1}-\beta}  \,,
\\
W'(r) &\sim  C_3 |S|^{\frac{\gamma_1+1}{\gamma_1-1}-\beta} 
+ C_4|S|^{\frac{\gamma_2+1}{\gamma_1-1}-\beta}  \,.
\end{align} 
These terms give stronger constraints than the pressure term above. In general we have that if $\gamma_2 \le \gamma_1$, then all of the terms are bounded without additional constraints on $\gamma_2$. However, if $\gamma_2 > \gamma_1$ then we must also have
\be 
\gamma_2 \le \beta(\gamma_1-1) -1 \,,
\ee
to ensure boundedness of all terms. 

\subsubsection*{2.B The marginal case \texorpdfstring{\(\gamma_1=1\)}{gamma\_1=1}}
The estimate~\eqref{eq:I_rp_asymp} excludes \(\gamma_1=1\). In this case the leading singularity
of the matter integral is logarithmic. To simplify the following, let us write $\epsilon\equiv r-r_p\to 0^+$. Using $x^{D-2}\simeq r_p^{D-2}$ in the singular region, one finds
\begin{align}
I(r)&=\int_{\infty}^{r}x^{D-2}\vrho(x)\,dx
\nonumber 
\\
&\sim A\,r_p^{D-2}\log \epsilon + C,
\qquad r\to r_p^+,
\label{eq:I_rp_marginal}
\end{align}
for some finite constant \(C\). Substituting into the definition of \(S(r)\) then gives
\begin{align}
S(r)&\sim S_0+\frac{2A}{r_p}\log\!\epsilon
\equiv S_0+\lambda_{\log}\log\!\epsilon,
\nonumber 
\\
\lambda_{\log} &\equiv \frac{2A}{r_p},
\qquad (r\to r_p^+),
\label{eq:S_rp_marginal}
\end{align}
so that \(|S|\to\infty\) as \(r\to r_p^+\), but only logarithmically. In particular,
\begin{equation}
|S|\sim |\lambda_{\log}|\,|\log\epsilon|,
\quad\Rightarrow\quad
\epsilon\sim e^{-|S|/|\lambda_{\log}|}.
\label{eq:eps_of_S_marginal}
\end{equation}
Consequently, the matter fields grow exponentially in \(|S|\):
\begin{align}
\vrho &\sim \frac{A}{\epsilon}\sim e^{|S|/|\lambda_{\log}|},
\quad
r\vrho'\sim \frac{1}{\epsilon^{2}}\sim e^{2|S|/|\lambda_{\log}|},
\nonumber 
\\
\p_r &\sim \epsilon^{-\gamma_2}\sim e^{\gamma_2|S|/|\lambda_{\log}|}.
\label{eq:scalings_rp_marginal}
\end{align}
Under the polynomial decay assumption~\eqref{eq:H_decay_beta}, these scalings are too strong to be
suppressed by \(H'(S)\) and \(H''(S)\). For example,
\begin{align}
\vrho H'(S) &\sim e^{|S|/|\lambda_{\log}|}\,|S|^{-\beta}\to\infty,
\nonumber 
\\
r\vrho' H'(S) &\sim e^{2|S|/|\lambda_{\log}|}\,|S|^{-\beta}\to\infty\,,
\end{align}
and similarly \(\vrho^2H''(S)\) and \(\vrho S H''(S)\) diverge. Thus, already in the case with
\(T^t_t=T^r_r\) (so that \(W= 0\)), the combinations listed in~\eqref{eq:dangerous_terms} are unbounded
for \(\gamma_1=1\), and the curvature invariants diverge. The anisotropic case only worsens this conclusion
through the additional contributions involving \(\p_r\) and \(W\).

\section{Faster-than-polynomial decays}
\label{app:fast_decay}

Appendix~\ref{app:blowups_unified} assumes the power-law decay~\eqref{eq:H_decay_beta}, which is
sufficient to bound the \emph{vacuum} curvature invariants and leads to the various exponent
inequalities on \(\gamma_1\) and \(\gamma_2\). Here we summarize what changes if instead the
resummation decays faster than any polynomial at large \(|S|\), in the sense that
\begin{align}
\lim_{|S|\to\infty}|S|^{n}H'(S)=0, \quad  \forall\,n>0,
\nonumber 
\\
\lim_{|S|\to\infty}|S|^{n}H''(S)=0, \quad  \forall\,n>0.
\label{eq:H_fast_decay}
\end{align}
A representative example is \(H'(S)\sim e^{-a|S|}\), or more generally \(H'(S)\sim e^{-a|S|^\sigma}\) with \(\sigma>0\). The results of this section are rather simple to obtain from the previous section---one must just track the various powers of $\beta$ that appear and replace them with the new super-polynomial decays. We therefore present a summary of the key facts. 

\subsection*{A. Power-law blow-ups that drive \texorpdfstring{\(|S|\to\infty\)}{|S|->infinity}}
In all power-law blow-ups treated in Appendix~\ref{app:blowups_unified} for which \(|S|\to\infty\)
and the matter fields grow at most polynomially in \(|S|\) (possibly with logarithmic corrections),
the faster-than-polynomial decay~\eqref{eq:H_fast_decay} implies that any factor of the form
\(|S|^{k}H'(S)\) or \(|S|^{k}H''(S)\) is bounded for every fixed \(k\ge 0\). Consequently, once
\(|S|\to\infty\), the density- and pressure-driven combinations in~\eqref{eq:dangerous_terms} are
automatically bounded, and the remaining \(W\)-dependent terms are likewise suppressed after using
that \(W\) carries an explicit factor of \(H'(S)\) through the field equations.

In particular, the exponent inequalities derived in Appendix~\ref{app:blowups_unified} become
\emph{sufficient but not necessary} in the following regimes:
\begin{itemize}
\item Central power-law blow-ups at \(r=0\) (including the marginal case \(\gamma_1=D-1\)):
once vacuum curvature is bounded, no additional restrictions on \(\gamma_1\) are needed, and the
pressure bounds on \(\gamma_2\) found in Case~1 correspond to sufficient conditions only.
\item Non-integrable finite-radius blow-ups with \(\gamma_1>1\):
the conditions such as~\eqref{eq:finite_r_condition} and the associated pressure and $W'$ bounds are
again sufficient but not necessary.
\end{itemize}
In these cases, vacuum boundedness implies boundedness in the presence of matter.

\subsection*{B. Cases where faster decay does not automatically help}
There remain two situations in which~\eqref{eq:H_fast_decay} does not by itself guarantee
regularity:
\begin{enumerate}
\item \textbf{Integrable finite-radius blow-up \(0<\gamma_1<1\).}
As discussed in Appendix~\ref{app:blowups_unified}, the integral \(I(r)\) stays finite and \(S(r)\)
does not run to \(\pm\infty\). In this case the large-\(|S|\) behaviour of \(H\) is irrelevant and
the curvature invariants still diverge.

\item \textbf{Marginal finite-radius blow-up \(\gamma_1=1\).}
From~\eqref{eq:S_rp_marginal}--\eqref{eq:eps_of_S_marginal} one has
\(S(r)\sim S_0+\lambda_{\log}\log\epsilon\) with \(\epsilon=r-r_p\to 0^+\), hence
\(\epsilon\sim e^{-|S|/|\lambda_{\log}|}\) and the matter fields grow exponentially in \(|S|\),
see~\eqref{eq:scalings_rp_marginal}. In this case the outcome depends on the rate of decay
of \(H'(S)\) and \(H''(S)\), not just that it is faster than polynomial. For example, if
\(H'(S)\sim e^{-a|S|}\) (and similarly for \(H''\)), then
\be
H'(S)\sim \epsilon^{a|\lambda_{\log}|},
\ee
and the most singular density-driven terms behave as
\begin{align}
r\vrho'\,H'(S) &\sim \epsilon^{-2}\,\epsilon^{a|\lambda_{\log}|}
=\epsilon^{a|\lambda_{\log}|-2},
\\
\vrho^{2}\,H''(S) &\sim \epsilon^{-2}\,\epsilon^{a|\lambda_{\log}|}
=\epsilon^{a|\lambda_{\log}|-2}.
\end{align}
Thus a sufficient condition in the $T_t^t = T_r^r$ case is \(a|\lambda_{\log}|\ge 2\). In the anisotropic case (at finite radius $r_p$ or at the inner horizon $r_h$), one also needs to control terms involving \(\p_r\) and \(W'(r)\); a simple sufficient condition is
\be
a|\lambda_{\log}|\ge \max\{2,\gamma_2+1\}.
\ee
For stronger decays such as \(H'(S)\sim e^{-a|S|^\sigma}\) with \(\sigma>1\), the suppression is stronger still and the marginal case is correspondingly easier to regularise.
\end{enumerate}

\section{Cosmologies}\label{app:OS}

In this appendix we remove the staticity condition and consider the fate of curvature invariants for cosmological spacetimes filled with a combination of perfect fluids.

The Friedmann-Lema\^itre-Robertson-Walker (FLRW) metric is given by
\begin{equation} \label{eq:dsFLRW}
\dif s^2 = -\dif t^2 + a(t)^2\left(\frac{\dif r^2}{1-kr^2}+r^2\dif\Omega^2_{D-2}\right)
\end{equation}
where $a(t)$ is the scale factor and $k=0,\pm1$ the spatial curvature. As argued in~\cite{Bueno:2025gjg}, spherical FLRW spacetimes can be easily studied within the two-dimensional Horndeski framework which captures the dynamical information of the spherically symmetric sector of QT theories. In particular, the scale factor satisfies second-order equations of motion for such theories.

Due to the isotropy of the spacetime metric, matter can be described by a perfect fluid, so that the stress-tensor takes the form~\eqref{eq:Tab} with $p_r=p_t\equiv p$. We then assume the presence of $n$ different matter sources, each characterized by an energy density $\rho_i$ and pressure $p_i$. The conservation law for each stress-energy tensor provides the following system of equations
\begin{equation}
    \dot{\varrho}_i + (D-1)\big(\varrho_i+\p_i\big)\frac{\dot{a}(t)}{a(t)} = 0\,,
\end{equation}
where the dot refers to the derivative with respect to $t$. The modified Friedmann equation then reads
\begin{equation}
    h(\Phi) = S(t) \,,
\end{equation}
where
\begin{equation}
     \quad \Phi=\frac{k+\dot{a}(t)^2}{a(t)^2}\,,\quad S(t)= \frac{2}{D-1} \sum_{i=1}^{n} \varrho_i(t)\,,
\end{equation}
where $h(x)$ is the characteristic function of QT gravities defined in~\eqref{char_poly}.

As argued in Sec.~\ref{sssm}, to determine the conditions on the matter for bounded curvature invariants it is necessary to analyse each term of the Riemann tensor, which in this case can be written as
\begin{equation}
R_{ab}^{\phantom{ab}cd} = \mathcal{R}^{(1)}\tau_{[a}^{[c} \pi_{b]}^{d]} + \mathcal{R}^{(2)} \pi_{[a}^{[c} \pi_{b]}^{d]} \,,
\end{equation}
where we have defined
\begin{equation} \label{eq:Ricosmo}
\mathcal{R}^{(1)} = 4\,\frac{\ddot{a}(t)}{a(t)} \,, \quad \mathcal{R}^{(2)} = 2\,\frac{k+\dot{a}(t)^2}{a(t)^2} \,.
\end{equation}
Here $\tau_a^b$ and $\pi_a^b$ are projectors onto the $t$ and spatial directions, respectively, defined in an analogous way to those that appear in~\eqref{eq:RformNf}. The Kretschmann scalar reads
\begin{align}\nonumber
    R_{abcd}R^{abcd} &= \frac{D-1}{4} \big(\mathcal{R}^{(1)}\big)^2 
    \\ & + \frac{(D-1)(D-2)}{2} \big(\mathcal{R}^{(2)}\big)^2 \,.
\end{align}

Using the Friedmann equation together with the conservation law of the stress-energy tensor, the $\mathcal{R}^{(i)}$ terms~\eqref{eq:Ricosmo} can be expressed in terms of the inverse of the characteristic function $H(x)=h^{-1}(x)$ as
\begin{align}
    \mathcal{R}^{(1)} &= 4 H(S) - 4 \sum_{i=1}^{n} (\varrho_i + \p_i) H'(S) \,,
    \\
    \mathcal{R}^{(2)} &= 2 H(S) \,.
\end{align}
Hence, in order to determine the validity of the limiting curvature hypothesis for this type of spacetimes we have to study the boundedness of the following quantities
\begin{equation}
    \{H(S)\,,\, S H'(S)\,,\, \p_i H'(S)\}\,.
\end{equation}
The first two are bounded by the analogous discussion in the static case. In order to study the third one, we assume that each pressure can be described by a polytropic equation of state with polytropic exponent $\Gamma_i\geq1$ at least for larger values of $|S|$, that is
\begin{equation}
    |\p_i| \sim |\varrho_i|^{\Gamma_i} \,.
\end{equation}
The sum of the pressures in $\mathcal{R}^{(1)}$ will then be dominated by the $j$th fluid with exponent $\Gamma_j=\max\{\Gamma_i:i=1,\hdots,n\}$ in the limit $|S|\to\infty$
\begin{align}
    \left\vert\sum_{i=1}^{n} \p_i\right\vert \sim |\p_j| \sim |\varrho_j|^{\Gamma_j} \sim |S|^{\Gamma_j} \,.
\end{align}
Therefore, since $H'(S)\sim|S|^{-\beta}$ with $\beta>1$, curvature remains bounded if and only if $\Gamma_i\leq\beta$ for all species of matter.

It is worth mentioning that for the usual cosmological equation of state $\p=w\varrho$, corresponding to $\Gamma=1$, the curvature of the cosmological solution is bounded provided that the curvature of the vacuum solution of the theory is bounded.

\subsection{Oppenheimer-Snyder collapse}
In a previous paper we studied the problem of gravitational collapse for a spherically symmetric perfect-fluid pressureless star---analogous to the original setup of Oppenheimer and Snyder~\cite{PhysRev.56.455}---in QT theories whose vacuum solutions are regular black holes~\cite{Bueno:2025gjg}. We showed that the stars undergo a never ending process of contractions and expansions throughout infinite universes and produce regular black holes as byproducts.

Since this process consists of matching the vacuum metric to the dust cosmological one at the star's boundary, it is straightforward that the limiting curvature hypothesis is satisfied during the collapse because both spacetimes satisfy it independently. 

Nonetheless, another question remains: is the global upper curvature bound given by the vacuum solution or by the matter solution? In order to address this question we should compare both Riemann tensors, corresponding to the interior and the exterior, during the collapse process. 

\begin{figure}[t!]
    \centering
    \includegraphics[width=0.48\textwidth]{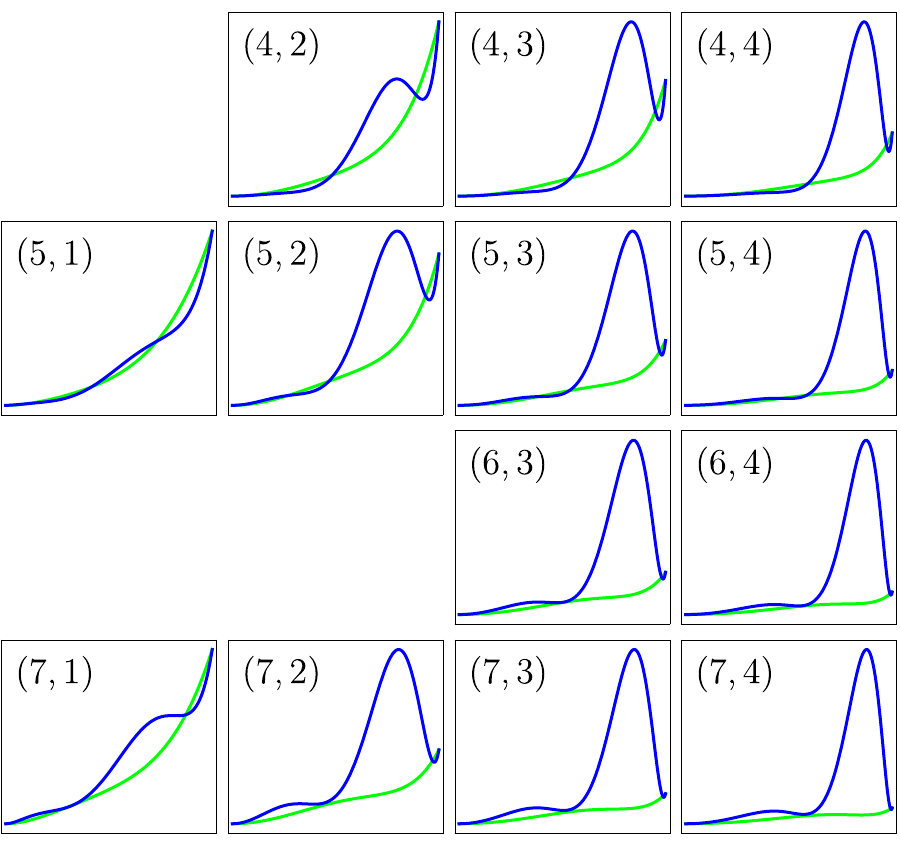}
    \caption{
    Plots of the Kretschmann scalar $\alpha^2R_{abcd}R^{abcd}$ for various $(D,\N)$-Hayward-like QT theories. In each graph, the curvature invariant is depicted in blue for the vacuum spacetime and in green for the FLRW spacetime, as a function of $\alpha\psi$ or $\alpha\Phi$, respectively, over their domain $(0,1)$. In all the cases presented, the maximum curvature is reached by the vacuum solution and it becomes more pronounced as $D$ and/or $\N$ increase.
    }
    \label{fig_OS}
\end{figure}

Making use of the scaling property in the vacuum and the homogeneity of the cosmological solutions, and expressing the couplings $\alpha_n=\alpha^{n-1}\beta_{n}$ for a given squared-length scale $\alpha>0$, both dimensionless Riemann tensors $\alpha R_{abcd}$---namely, the one corresponding to the vacuum metric as well as the one corresponding to the FLRW metric---can be expressed as functions of $\alpha\psi$ and $\alpha\Phi$ alone. Given that both quantities depend in a complicated way on the characteristic polynomial, let us consider Hayward solutions in order to analyse whether the maximum value of the Kretschmann scalar during a collapse process corresponds to a fixed value given by the vacuum theory. Since the collapse depends on the initial conditions, we propose an alternative approach: comparing the two curvature invariants separately within their corresponding domains for Hayward solutions. This problem cannot be tackled analytically, so we address it graphically. In Fig.~\ref{fig_OS} we plot both curvature invariants for different dimensions $D$ and orders $\N$ for Hayward-like solutions. We observe that the Kretschmann scalar in the FLRW spacetime always lies below its maximum value in vacuum. Furthermore, for higher $D$ and/or $\N$, the maximum curvature in vacuum becomes significantly larger than in FLRW. We thus conclude that Markov's hypothesis holds in Hayward-like cosmologies with dust, and therefore also in this collapsing scenario.

\bibliographystyle{JHEP-2}
\bibliography{Gravities}
\noindent 

\end{document}